\documentclass[11pt]{article}
\usepackage{osid}
\usepackage{graphicx} % Required for inserting images

\usepackage{braket, mathtools, amssymb, bbm, amsmath, mathrsfs, bm, hyperref, color, amsthm, xspace, bm, comment}

\usepackage[style=phys,articletitle=false,biblabel=brackets,chaptertitle=false,pageranges=false]{biblatex}
\newcommand{\ketbra}[1]{\ket{#1}\bra{#1}}
\newcommand{\id}{\mathbbm{1}}
\newcommand{\abs}[1]{\lvert #1 \rvert}
\newcommand{\norm}[1]{\lVert #1 \rVert}
\newcommand{\tr}{\operatorname{tr}}
\newcommand{\im}{\operatorname{Im}}

\newcommand{\schro}{Shr\"odinger\xspace}

\title{Quantum jump unravelings for non-Markovian open system dynamics: a review}

\author{Federico Settimo and Jyrki Piilo\\
\it
Department of Physics and Astronomy, University of Turku, FI-20014 Turun yliopisto, Finland\\
e-mail: fesett@utu.fi, jyrki.piilo@utu.fi}
\begin{document}

\maketitle
\begin{abstract}
    Stochastic unravelings provide a useful way to represent open quantum system dynamics in terms of pure state realizations, and have been widely studied both from a fundamental and from a computational point of view.
    They were initially formulated for Markovian dynamics described by the Gorini-Kossakowski-Sudarshan-Lindblad master equation.
    However, due to recent technological and experimental development, most physical relevant dynamics present temporal correlations beyond the Markov approximation.
    Such correlations cause decay rates to turn temporarily negative, thus requiring the generalization of stochastic unravelings from Markovian to non-Markovian scenarios.
    Indeed, many unraveling techniques have been introduced in this regime, and a comprehensive review of the different jump methods is currently missing.
    In this work, we provide an overview of widely used quantum jump unraveling techniques for non-Markovian systems and also discuss them in terms of their numerical efficiency, divisibility requirements, Hilbert space extension, and measurement interpretation.
\end{abstract}

%\tableofcontents

% --------------------------------------- Intro ---------------------------------------
\section{Introduction}
\label{sec:intro}
For Markovian dynamics described by the Gorini-Kossakowski-Sudarshan-Lindblad (GKSL) master equation \cite{Gorini1976, Lindblad1976}, stochastic unravelings provide an equivalent representation of the reduced dynamics in terms of pure state realizations.
This fact allows for a powerful tool to simulate the time evolution of open quantum systems \cite{Breuer-Petruccione, Rivas-Huelga-OQS, Vacchini-OQS}.
The main idea behind unravelings is to describe the dynamics of a $d\times d$ density matrix as the stochastic average of $d$-dimensional pure state vectors, where $d$ is the dimension of the Hilbert space, thus drastically improving the simulation requirements.
They fall into two major classes, corresponding to physically different continuous monitoring schemes of the environment.
The resulting stochastic processes can thus be in two forms: they can be either diffusive \cite{Gisin1992, Percival1999} or piecewise deterministic \cite{Dalibard1992Wave-functionOptics, Dum1992, Mlmer1993MonteOptics, Plenio1998}, in which the deterministic evolution is interrupted by random discontinuous jumps.

The importance of stochastic unravelings is not limited to their numerical efficiency.
In fact, in the Markovian regime they can be interpreted as a continuous measurement acting on the system \cite{Barchielli1991, Wiseman2009QuantumControl, Gardiner2004QuantumOptics, Barchielli2009, Albarelli2024}.
Such continuous measurement techniques allow for improved control \cite{Wiseman1993, Wiseman1994, Zhang2017}, error correction \cite{Borah2022, Livingston2022} and parameter estimations \cite{Gammelmark2013, Gammelmark2014FisherMeasurements, Kiilerich2016, Guevara2020, Bao2020, Radaelli2026}, and they have been employed in several experimental platforms, including superconducting qubits \cite{Gambetta2008, Riste2013, Roch2014, Campagne-Ibarcq2016, Weber2016, Ficheux2018, Minev2019}, trapped ions \cite{Bergquist1986, Viola1997, Leibfried2003, Ozawa2018}, cavity QED \cite{Smith2002, Gleyzes2007, Sayrin2011,Rybarczyk2015} and molecular compounds \cite{Basche1995}.

Nevertheless, the Markov approximation only holds for systems weakly coupled with their environments and in general does not hold for realistic open systems.
Violations of the Markov approximation have been widely studied and connected to memory effects and information backflows \cite{BLPV-colloquim, rivas-quantum-nm, Buscemi2016, Li2018ConceptsHierarchy, DeVega2017, Chruscinski2022}.
Different inequivalent definitions of quantum non-Markovianity have been introduced and connected to revivals in the distinguishability between quantum states \cite{BLP-PRL, BLP-PRA, Chruscinski2011, Wimann2012OptimalDynamics, Smirne2022, Settimo-JSD} and entanglement \cite{RHP}.
Such non-Markovian effects have also been demonstrated in numerous experimental platforms \cite{Liu2011, Liu2013, Liu2018, Cialdi2019, White2020, Goswami2021}.

Crucially for stochastic unravelings, the presence of non-Markovianity implies the negativity of some of the rates in the GKSL master equation \cite{Kossakowski1972OnSemigroup, Hall2014}.
This fact raises both conceptual and computational challenges, since the Markovian unraveling techniques would predict unphysical negative probabilities, and therefore cannot be employed to simulate the dynamics.
Therefore, the simulation techniques become more expensive if one wants to take into account the realistic memory effects present in the dynamics, as well as challenging the interpretation of trajectories as arising from continuous monitoring of the system alone.

Nevertheless, in the recent years numerous simulation techniques have been introduced, for a recent review of deterministic approaches see \cite{Xu2026Review}.
On the other hand, also numerous stochastic unraveling techniques have been introduced, and no unified comparison of the different techniques exists in the literature.
In this work, we discuss several jump unraveling techniques which allow to efficiently simulate non-Markovian open quantum system dynamics.
Notice that unraveling techniques to describe non-Markovian dynamics have been introduced also in the diffusive case \cite{Diosi-NMQSD, Caiaffa-W-diffusive, Luoma-diffusive-NMQJ}.

The rest of the work is organized as follows.
In Sec.~\ref{sec:OQS}, we provide a brief overview of open quantum system dynamics, quantum non-Markovianity and some widely used stochastic unraveling techniques for the Markovian case.
In Sec.~\ref{sec:Hilbert_space}, we present some jump unravelings for non-Markovian dynamics that do not require the extension of the system's Hilbert space.
In Sec.~\ref{sec:extended_Hilbert}, instead, we present techniques that do require the extension of the Hilbert space.
Then, in Sec.~\ref{sec:examples}, we compare the unraveling methods on a number of common physically relevant examples.
In Sec.~\ref{sec:beyond_Lindblad}, we address extensions to scenarios beyond the non-Markovian GKSL master equation.
Then, in Sec.~\ref{sec:measurement}, we briefly discuss when unravelings can be seen as arising from continously measuring the open system.
Lastly, in Sec.~\ref{sec:discussion}, we present a summary and overlook of this review.

% --------------------------------------- OQS ---------------------------------------
\section{Background: open quantum systems}
\label{sec:OQS}

In this section, we provide a brief introduction to open quantum systems.
We first consider systems weakly interacting with their environments, so that the dynamics is described by the GKSL master equation.
We then describe some widely used jump-like unraveling techniques to describe the dynamics in such regime.
This is followed by a short introduction to realistic open quantum systems, in which the interaction does not need to be weak, and the dynamics can present memory effects, causing a failure of the Markovian unraveling techniques.

\subsection{GKSL master equation}
\label{subsec:Lindblad}
Realistic open quantum systems are unavoidably coupled to external and uncontrollable degrees of freedom, the so called environment.
Under the assumption that the system and the environment interact weakly and are initially uncorrelated, then the reduced system dynamics is well described by the GKSL master equation \cite{Gorini1976, Lindblad1976}
\begin{equation}
    \label{eq:Lindblad_semigroup}
    \frac d{dt}\rho =\mathcal L[\rho] =-i[H,\rho]+ \sum_\alpha \gamma_\alpha L_\alpha\rho L_\alpha^\dagger-\frac12\{\Gamma,\rho\},
\end{equation}
where we set $\hbar=1$.
The rates $\gamma_\alpha\ge0$ are non-negative numbers and
\begin{equation}
    \label{eq:Gamma}
    \Gamma =  \sum_\alpha \gamma_\alpha L_\alpha^\dagger L_\alpha.
\end{equation}
Above, $L_\alpha$ are called jump operators.
The solution is given by $\rho(t) = \Lambda_t[\rho(0)]$, with $\Lambda_t$ being completely positive (CP) and trace preserving (TP).
Formally, the dynamical map can be written as
\begin{equation}
    \label{eq:exp_L_semigroup}
    \Lambda_t = \exp\{\mathcal L\, t\},
\end{equation}
although in general there is no closed form for the exponential, and it obeys the semigroup property \cite{Alicki2007}
\begin{equation}
    \label{eq:semigroup}
    \Lambda_{t+s} = \Lambda_t\Lambda_s = \Lambda_s\Lambda_t
\end{equation}
for all $t,s\ge0$.

\subsection{Markovian jump-like Monte Carlo methods}
\label{subsec:MC_Markov}
The GKSL master equation \eqref{eq:Lindblad_semigroup}, although having a very simple and appealing form, is in general very difficult to solve analytically and therefore one usually resorts to numerical methods to simulate its solution.
However, na\"ive numerical methods can be computationally very expensive to simulate when dealing with high-dimensional systems, the main reason being that all operators in Eq.~\eqref{eq:Lindblad_semigroup} are $d\times d$ matrices, with $d$ being the dimension of the Hilbert space.
Stochastic unravelings are a widely used tool to solve this problem.

They are based on the fact that, due to the convexity of the set of quantum states, the solution of the GKSL equation at a given time $t$ can always be written as the expectation value of a stochastic process on the set of pure states
\begin{equation}
    \label{eq:avg_solution}
    \rho(t) = \mathbb E\left[\ketbra{\psi(t)}\right],
\end{equation}
where $\mathbb E[\cdot]$ represents the ensemble average
\begin{equation}
    \mathbb E\left[\ketbra{\psi(t)}\right] \coloneqq \int \mathcal D\psi\ p(\psi,t)\ \ketbra{\psi(t)},
\end{equation}
where each pure state $\ket{\psi(t)}$ in the ensemble evolves according to a stochastic evolution, and $p(\psi,t)$ is a probability density on the set of pure states
\begin{equation}
    p(\psi,t)\ge0,\qquad\int\mathcal D\psi\ p(\psi,t) = 1.
\end{equation}
Computationally, stochastic unravelings are performed by sampling a set of $N$ realizations of the stochastic process, from which the state is reconstructed as
\begin{equation}
    \label{eq:avg_solution_approx}
    \rho(t)\approx\frac1{N}\sum_i\ketbra{\psi_i(t)},
\end{equation}
where $\ket{\psi_i(t)}$ denotes the $i$-th realization.
Such stochastic sampling drastically improves the efficiency of the simulations, since it is sufficient to deal with $d$-dimensional pure states instead of $d\times d$ matrices.
The various unraveling methods differ depending on the specific stochastic process considered for the evolution of the state vector $\ket{\psi(t)}$ and on the eventual need to extend the Hilbert space.

In the following, we will only consider piecewise-deterministic processes, in which the evolution of the stochastic realization $\ket{\psi(t)}\mapsto\ket{\psi(t+dt)}$ in a small timestep $dt$ can either be continuous and deterministic or interrupted by infrequent discontinuous jumps, happening with probability proportional to the timestep $dt$.

All unraveling techniques, as typical for computational methods, agree with the exact solution up to the first order in $dt$, and therefore all terms of the order $dt^2$ or higher are neglected.
Nevertheless, it is possible to tune the stochastic unravelings to have precision to arbitrary order in $dt$ \cite{Steinbach1995High-orderEvolution}.

\subsubsection{Monte Carlo wave function}
\label{subsubsec:MCWF}

The simplest unraveling method is the so-called Monte Carlo wave function (MCWF) \cite{Dalibard1992Wave-functionOptics, Dum1992, Mlmer1993MonteOptics, Plenio1998}.
Suppose that, at a given time $t$, the stochastic realization is in the state $\ket{\psi(t)}$.
It can either evolve via a stochastic discontinuous jump
\begin{equation}
    \label{eq:MCWF_jump}
    %\ket{\psi(t)}\mapsto\ket{\psi_{\alpha}(t+dt)}\coloneqq \frac{L_\alpha\ket{\psi(t)}}{\norm{L_\alpha\ket{\psi(t)}}}
    \ket{\psi(t)}\mapsto\ket{\tilde\psi_{\alpha}^{\text{MC}}(t+dt)}\coloneqq {L_\alpha\ket{\psi(t)}}
\end{equation}
with probability
\begin{equation}
    \label{eq:MCWF_p_jump}
    p^{{\text{MC}}}_\alpha = \gamma_\alpha \norm{L_\alpha\ket{\psi(t)}}^2\ dt,
\end{equation}
or it can evolve deterministically
\begin{equation}
    \label{eq:MCWF_det}
    %\ket{\psi(t)}\mapsto\ket{\psi_{\text{det}}(t+dt)}\coloneqq \frac{\left(\id - i K dt\right)\ket{\psi(t)}}{\norm{\left(\id - i K dt\right)\ket{\psi(t)}}},
    \ket{\psi(t)}\mapsto\ket{\tilde\psi_{\text{det}}^{\text{MC}}(t+dt)}\coloneqq {\left(\id - i K dt\right)\ket{\psi(t)}},
\end{equation}
where $K$ is an effective non-Hermitian Hamiltonian depending on both the unitary and dissipative part of the master equation
\begin{equation}
    \label{eq:MCWF_K}
    K = H-\frac i2\Gamma.
\end{equation}
The corresponding probability of deterministic evolution is
\begin{equation}
    \label{eq:MCWF_p_det}
    p_{\text{det}}^{\text{MC}} = 1-\sum_\alpha p^{{\text{MC}}}_\alpha.
\end{equation}
The deterministic evolution can be equivalently written as the solution of the \schro equation
\begin{equation}
    \label{eq:MCWF_Schro_eq}
    \frac d{dt}\ket{\tilde\psi_{\text{det}}^{\text{MC}}(t)} = -i K\ket{\tilde\psi_{\text{det}}^{\text{MC}}(t)}.
\end{equation}

After the jump or deterministic evolution in the timestep $dt$, the state vector is normalized
\begin{equation}
    \ket{\tilde\psi(t+dt)}\mapsto\ket{\psi(t+dt)} = \frac{\ket{\tilde\psi(t+dt)}}{\norm{\ket{\tilde\psi(t+dt)}}},
\end{equation}
where $\ket{\tilde\psi(t+dt)}$ refers to either $\ket{\tilde\psi_\alpha^{\text{MC}}(t+dt)}$ or $\ket{\tilde\psi_{\text{det}}^{\text{MC}}(t+dt)}$.
In the following, we will always use a tilde to denote unnormalized state vectors $\ket{\tilde\phi}$, with the correspondent normalized vector given by $\ket{\phi} = \ket{\tilde\phi}/\norm{\ket{\tilde\phi}}$.

The average evolution in the timestep $dt$ is therefore given by
\begin{equation}
    \begin{split}
        \Big\langle\ketbra{\psi(t+dt)}\Big\rangle_{\text{MC}} =&\  p_{\text{det}}^{\text{MC}}\ketbra{\psi_{\text{det}}^{\text{MC}}(t+dt)}\\
        &+ \sum_\alpha p^{{\text{MC}}}_\alpha\ketbra{\psi_\alpha^{\text{MC}}(t+dt)}\\
        =&\ \ketbra{\psi(t)} +\mathcal L\left[\ketbra{\psi(t)}\right]\ dt,
    \end{split}
\end{equation}
where $\braket\cdot_{\text{MC}}$ is the expectation value over the stochastic process defined by the MCWF unraveling.
Therefore, if the solution at time $t$ is represented as in Eq.~\eqref{eq:avg_solution_approx}, then 
\begin{equation}
    \begin{split}
        \rho(t+dt) &= \frac1 N\sum_i\Big\langle\ketbra{\psi_i(t+dt)}\Big\rangle\\
        &=\frac1N\sum_i\Big( \ketbra{\psi_i(t)} +\mathcal L\left[\ketbra{\psi_i(t)}\right]\ dt\Big)\\
        &= \rho(t) + \mathcal L\left[\rho(t)\right]\ dt.
    \end{split}
\end{equation}

\begin{figure}
    \centering
    \includegraphics[width=0.8\linewidth]{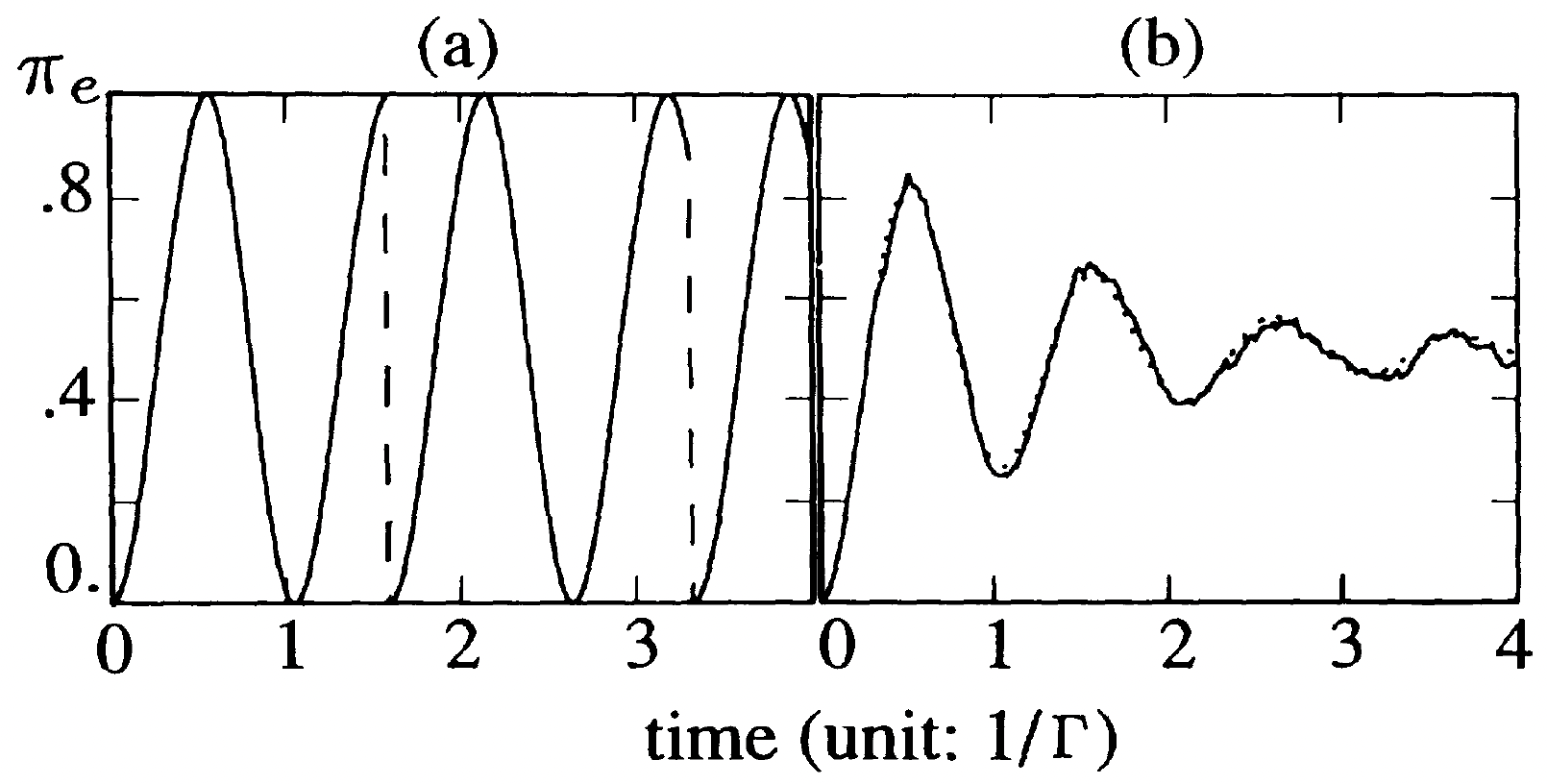}
    \caption{Figure from \cite{Dalibard1992Wave-functionOptics}.
    Panel (a): excited state population for a typical trajectory; after the detection of a photon the realization is in the ground state.
    Panel (b): average over 100 realizations.}
    \label{fig:MCWF}
\end{figure}

The stochastic realizations $\{\ket{\psi_i(t)} : 0\le t\le T\}$ are not only useful for the simulation of the average evolution, but can also be interpreted as the result of a continuous measurement process acting on the system.
To exemplify this view, let us consider the example of a two-level atom undergoing spontaneous emission.
Let $\ket 0$ and $\ket 1$ represent, respectively, the ground and excited state of the atom.
The system Hamiltonian is given by $H_{\text{em}}=\frac{\omega_0}2\ \sigma_z + \omega\ \sigma_x$, where $\sigma_{x,z}$ are the Pauli matrices, $\omega_0$ is the energy difference and $\omega$ the driving frequency.
The spontaneous emission is described by the dissipative part of the GKSL master equation \eqref{eq:Lindblad_semigroup} and the corresponding jump operator is $L=\sigma_- = \ket0\bra1$, with rate $\gamma$.
The jump process of Eq.~\eqref{eq:MCWF_jump} then becomes
\begin{equation}
    \ket{\psi(t)}\mapsto\ket0,\qquad p^{\text{MC}}_- = \gamma\ \abs{\braket{1\vert\psi(t)}}^2\ dt.
\end{equation}
This can be interpreted as the detection of a photon emitted by the atom in the timestep $dt$: after the emission process, the atom is found to be in its ground state $\ket0$ and the probability of such detection event is given by the Born rule.
If, instead, no emission is observed in the timestep $dt$, then the time evolution is governed by the effective Hamiltonian
\begin{equation}
    K_{\text{em}} = \frac{\omega_0}2 \sigma_z + \omega \sigma_x - \frac i2 \gamma \ketbra1.
\end{equation}
The continuous monitoring is affecting the dynamics even if no photon is recorded.
This change in the free dynamics is described by the non-Hermitian part of the Hamiltonian $- \frac i2 \gamma \ketbra1$.
For an illustration of the interplay between jump and deterministic evolution, as well as the continuous measurement interpretation, see Fig.~\ref{fig:MCWF}.

\subsubsection{Waiting time distribution}
\label{subsubsec:WTD}
The MCWF can be equivalently reformulated in terms of the waiting time distribution (WTD) of detection events \cite{Gardiner1992}.
Given an initial time $t_0$ and a state $\ket{\psi(t_0})$, the time evolution proceeds as following:
\begin{enumerate}
    \item A random number $x$ is drawn from the uniform distribution in $[0,1]$.
    \item The state evolves as
    \begin{equation}
        \label{eq:WTD_det}
        \ket{\tilde\psi_{\text{det}}^{\text{WT}}(t)} = e^{-i K(t-t_0)}\ket{\psi(t_0)},
    \end{equation}
    with $K$ as in Eq.~\eqref{eq:MCWF_K}.
    Notice that such state is the same as the one obtained via Eq.~\eqref{eq:MCWF_det} for the MCWF.
    \item The time $t_1$ of the jump is determined by the condition
    \begin{equation}
        \label{eq:WTD_jump_time}
        x = \braket{\tilde\psi_{\text{det}}^{\text{WT}}(t_1)\vert\tilde\psi_{\text{det}}^{\text{WT}}(t_1)}.
    \end{equation}
    \item The state for $t\in[t_0,t_1]$ is $\ket{{\psi}_{\text{det}}^{\text{WT}}(t)} = \ket{\tilde{\psi}_{\text{det}}^{\text{WT}}(t)}/\norm{\ket{\tilde{\psi}_{\text{det}}^{\text{WT}}(t)}}$.
    \item The state right after time $t_1$ is chosen randomly among the states
    \begin{equation}
        \ket{\psi_\alpha^{\text{WT}}(t_1)} \coloneqq \frac{L_\alpha\ket{\psi(t_1)}}{\norm{L_\alpha\ket{\psi(t_1)}}}
    \end{equation}
    with respective probability
    \begin{equation}
        \label{eq:WTD_p_jump}
        p^{\text{WT}}_\alpha(t_1) = \frac{\gamma_\alpha\braket{\psi_{\text{det}}^{\text{WT}}(t_1)\vert L_\alpha^\dagger L_\alpha\vert\psi_{\text{det}}^{\text{WT}}(t_1)}}{\braket{\psi_{\text{det}}^{\text{WT}}(t_1)\vert \Gamma\vert\psi_{\text{det}}^{\text{WT}}(t_1)}}.
    \end{equation}
    \item The process is repeated with $\ket{\psi_\alpha^{\text{WT}}(t_1)}$ as the initial state.
\end{enumerate}
The resulting stochastic trajectories are the same as the ones obtained with the MCWF procedure: the jumps are obtained by applying the operators $L_\alpha$ and the state $\ket{\tilde\psi_{\text{det}}^{\text{WT}}(t)}$ of Eq.~\eqref{eq:WTD_det} obeys the same \schro equation \eqref{eq:MCWF_Schro_eq}.

The use of the WTD distribution instead of computing the jump probability at each time step, is however not limited to the MCWF technique.
Indeed, WTD distributions can be defined also for different unraveling schemes, both Markovian and non-Markovian \cite{Luoma2012, Radaelli2024GillespieTrajectories}.

\subsection{Non-Markovian open quantum systems}
\label{subsec:nM_OQS}

The GKSL master equation \eqref{eq:Lindblad_semigroup} and the semigroup property \eqref{eq:semigroup} is a good description of the reduced dynamics only in the limit of system and environment weakly coupled.
However, this assumption is generally not physically valid, and therefore the unraveling schemes might no longer be applicable.

\subsubsection{Time-local master equation}
\label{subsubsec:NM_ME}
Under suitable regularity conditions, i.e. assuming that the dynamical map is invertible, then the reduced dynamics for realistic systems can be written in a GKSL form but with a time-dependent generator  \cite{Breuer-Petruccione, Rivas-Huelga-OQS, Vacchini-OQS}
\begin{equation}
    \label{eq:Lindblad}
    \mathcal L_t[\rho] =-i[H(t),\rho]+ \sum_\alpha \gamma_\alpha(t) L_\alpha(t)\rho L_\alpha^\dagger(t)-\frac12\{\Gamma(t),\rho\},
\end{equation}
with $\Gamma(t)$ as in Eq.~\eqref{eq:Gamma} and the rates $\gamma_\alpha(t)$ that can become temporarily negative.
The solution can be formally written as
\begin{equation}
    \Lambda_t = T \exp\int_0^td\tau\, \mathcal L_\tau,
\end{equation}
with $T\exp$ being the chronologically time ordered exponential, and $\Lambda_t$ is again a CPTP map.

As long as the rates are positive at all times, then both the MCWF and the WTD unravelings can be applied, up to considering the time-dependent rates and operators.
If, instead, the rates are negative, both methods fail since Eqs.~\eqref{eq:MCWF_p_jump} and \eqref{eq:WTD_p_jump} give negative probabilities, which is not feasible directly.
In the next sections, we will present some unraveling schemes that can be applied even in the presence of non-positive rates.

\subsubsection{Divisibility}
\label{subsubsec:divisibility}
For time-dependent rates, the semigroup condition \eqref{eq:semigroup} no longer holds.
However, it is possible to derive the dynamical map at time $t$ from the map at time $s<t$ as
\begin{equation}
    \label{eq:divisibility}
    \Lambda_t = V_{t,s}\Lambda_s,\qquad V_{t,s} = \Lambda_t\Lambda_s^{-1}.
\end{equation}
Although both $\Lambda_t$ and $\Lambda_s$ are CP, the propagator $V_{t,s}$ is not guaranteed to be CP: the dynamics is said to be (C)P divisible if $V_{t,s}$ is (C)P.
CP divisibility is equivalent to the non-negativity of all the rates $\gamma_\alpha(t)\ge0$, while it is P divisible if and only if the weaker condition \cite{Kossakowski1972OnSemigroup}
\begin{equation}
    \label{eq:P_divisibility}
    \sum_\alpha\gamma_\alpha(t)\abs{\braket{\varphi_\mu\vert L_\alpha(t)\vert\varphi_{\mu^\prime}}}^2\ge0
\end{equation}
holds for all orthonormal bases $\{\ket{\varphi_\mu}\}_\mu$ and for all $\mu\ne\mu^\prime$.

Divisibility has been widely studied in open system dynamics, and lack of it has been connected to memory effects and non-Markovianity \cite{BLP-PRA, BLP-PRL, RHP, rivas-quantum-nm, Wimann2015, BLPV-colloquim, Budini2022}.
From the point of view of unravelings, instead, the MCWF and WTD methods hold if and only if the dynamics is CP divisible, since it is the only case in which the jump probabilities are positive.

% --------------------------------------- Modes ---------------------------------------
\subsection{Pseudomodes}
\label{subsec:modes}

One possible way of applying the MCWF to non-Markovian systems is by embedding the environment into a set of finite auxiliary modes, known as pseudomodes \cite{Garraway1997DecayReservoir, Garraway1997NonperturbativeCavity, Dalton2001TheoryProcesses} or fictitious modes \cite{Imamoglu1994}.
Each of the resulting auxiliary mode will then interact with a residual Markovian bath.
Therefore, the Markovian Monte Carlo methods can be applied on the joint system + auxiliary modes.
This approach, although allowing for the application of Markovian methods, is limited because the unravelings must be on the enlarged system, thus increasing the numerical requirements, and the resulting stochastic dynamics on the reduced system does not consist in general of pure states.

%\subsubsection{Pseudomodes}
%\label{subsubsec:pseudomodes}
In the pseudomode approach, the system is assumed to be a $n$-levels atom interacting with a bosonic environment, with a global Hamiltonian of the form 
\begin{equation}
    H = \sum_i\omega_i \ketbra i + \sum_\lambda\omega_\lambda a^\dagger_\lambda a_\lambda + \sum_{i,\lambda}g^{(i)}_\lambda\left(\sigma_-^{(i)}\otimes a^\dagger_\lambda + \sigma_+^{(i)} \otimes a_\lambda\right),
\end{equation}
where the first term is the system free Hamiltonian, the second is the bosonic environment, with $a_\lambda^\dagger$ and $a_\lambda$ bosonic creation and annihilation operators.
The last term is the interaction term, with $\sigma_-^{(i)}=\ket0\bra i = (\sigma_+^{(i)})^\dagger$.

The resulting dynamics can be rewritten in terms of the pseudomodes creation and annihilation operators $\hat a_l^\dagger$ and $\hat a_l$ in GKSL form, with non-Hermitian Hamiltonian
\begin{equation}
    K_{\text{pm}} = \sum_i\omega_i \ketbra i + \sum_l z_l\, \hat a^\dagger_l \hat a_l + \sum_{i,l}g_{i,l}\left(\sigma_-^{(i)}\otimes \hat a^\dagger_l + \sigma_+^{(i)} \otimes \hat a_l\right).
\end{equation}
The non-Hermiticty arises from the fact that $z_l$ can have a non-zero (and negative) imaginary part.
The corresponding Lindblad operators are the pseudomode operators $L_l = \hat a_l$, and the corresponding rates are given by $\gamma_l = \sqrt{-2\im z_l}$.
The explicit form of pseudomodes operators and couplings can be 
found e.g.~in \cite{Garraway1997NonperturbativeCavity}.
These pseudomodes are physically motivated since the complex parameters $z_l$ correspond to poles of the environmental spectral density, and therefore the pseudomodes are related to resonances of the environment.

%\subsubsection{Fictitious modes}
%\label{subsubsec:fictitiousmodes}
An alternative approach is that of the so-called fictitious modes \cite{Imamoglu1994}, in which the resulting GKSL master equation has similar jump operators defined by the fictitious mode creation and annihilation operators.
Unlike the pseudomodes, the fictitious modes do not arise from physical properties of the environment but are only a mathematical construct.

% --------------------------------------- No extension Hilbert space ---------------------------------------
\section{Stochastic processes on the Hilbert space}
\label{sec:Hilbert_space}
In this section, we provide a description of widely used unraveling techniques that can allow to simulate non-Markovian master equation which do not require any extension of the Hilbert space.
There are two ways to circumvent the problem of the negativity of the rates: it is possible to either introduce correlations between the stochastic realizations or to define different jump processes, while the two are not mutually exclusive.
%The two approaches are not mutually exclusive: it is possible to have correlated realizations also for jump processes not described by the MCWF formalism.

%The modified jump processes require the definition of appropriate rate operators (RO) and the jumps occur to the corresponding eigenstates.
%Therefore, these methods are known as RO quantum jumps (ROQJ).

\subsection{Non-Markovian quantum jumps}
\label{subsec:NMQJ}
We now introduce the Non-Markovian quantum jump (NMQJ) technique \cite{Piilo2008, Piilo2009OpenJumps, Breuer2008}, which allows to deal with temporarily negative rates by introducing correlations among the different stochastic realizations and by allowing for a reverse-jump process which has the effect of canceling a jump that had previously happened.

Suppose that at a given time $t$ the dynamics is described by the effective ensemble $\{(N_i(t), \ket{\psi_i(t)})_i\}$, i.e.~there are $N_i(t)$ copies of the state $\ket{\psi_i(t)}$ such that
\begin{equation}
    \rho(t) = \sum_i\frac{N_i(t)}N\ketbra{\psi_i(t)},
\end{equation}
with $\sum_iN_i(t) = N$.
If the rate $\gamma_\alpha(t)$ is positive, then the jump process for each state $\ket{\psi_i(t)}$ is the same as for the MCWF $\ket{\psi_i(t)}\mapsto L_\alpha(t)\ket{\psi_i(t)}$ with probabilities as in Eq.~\eqref{eq:MCWF_p_jump}.
If, instead, the rate is negative $\gamma_\alpha(t)<0$, then the state can evolve via a reverse jump
\begin{equation}
    \label{eq:NMQJ_jump}
    \ket{\psi_i(t)} = \frac{L_\alpha(t)\ket{\psi_j(t)}}{\norm{L_\alpha(t)\ket{\psi_j(t)}}}\mapsto\ket{\psi_j(t)}
\end{equation}
happening with probability
\begin{equation}
    \label{eq:NMQJ_p_jump}
    p^{\text{NMQJ}}_{i\to j} = -\frac{N_j(t)}{N_i(t)}\gamma_\alpha(t)\norm{L_\alpha(t)\ket{\psi_j(t)}}^2\ dt.
\end{equation}
%This reverse jump process is possible only if the state $\ket{\psi_i(t)}$ is the target of the corresponding normal jump if the rates were positive and it has the effect of canceling the direct jump.
In other words, this reverse jump cancels the earlier occurred positive rate direct jump. 
The corresponding probability depends on both the current and the target state, thus the different realizations are not independent.
This fact, although allowing for unraveling of dynamics with negative rates, makes the simulations more expensive since one needs to keep track of all realizations at the same time.

\begin{figure}
    \centering
    \includegraphics[width=0.7\linewidth]{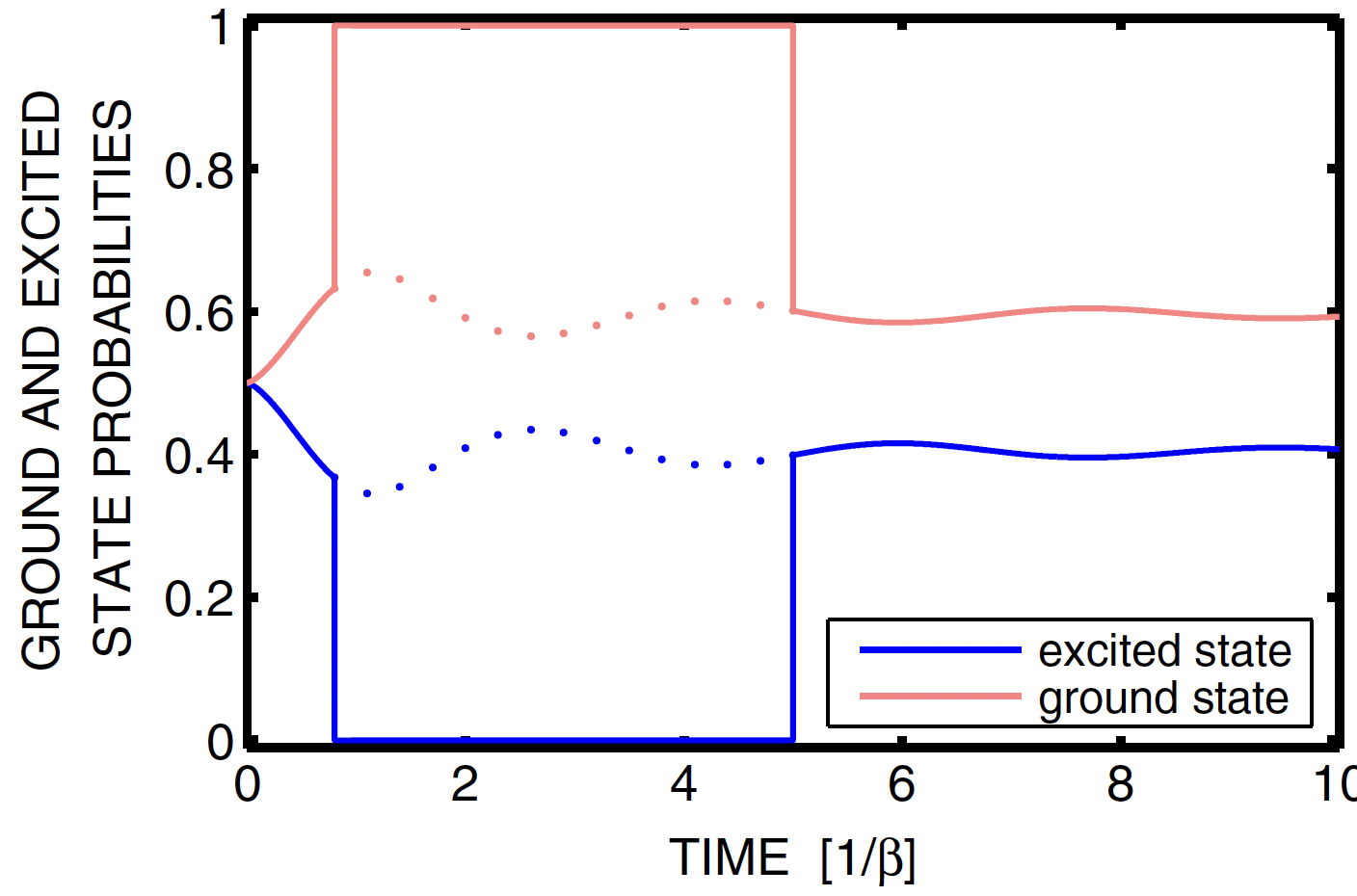}
    \caption{Figure from \cite{Piilo2008}, showing example realization undergoing a reverse jump: the reverse jump happening at time $\approx5\beta$ undoes the previous jump in the sense that the state after the reverse jump is the state in which the realization would have been in, if the direct jump at time $\approx0.8\beta$ would not have happened.}
    \label{fig:NMQJ}
\end{figure}

In Fig.~\ref{fig:NMQJ}, we present a realization of the reverse jumps: after the initial direct jump, the system is in $\ket0$ and the superposition is destroyed.
The reverse jump undoes the direct jump and recreates the superposition, by reverting the state to what would have been if the previous direct jump would not have happened.
This recreation of superposition can be viewed as restoring the information and it paved the way to the idea of information backflow in non-Markovian dynamics \cite{BLP-PRL, BLP-PRA, BLPV-colloquim}.

Notice that all other unravelings defined in the following of this section can be equipped with the reverse jumps of the NMQJ technique.
The jump has the action of canceling a direct jump, while the corresponding probability reads
\begin{equation}
    \label{eq:NMQJ_p_jump_ROs}
    p^{\text{NMQJ}, \star}_{i\to j} = -\frac{N_j(t)}{N_i(t)}\ p^{\star}_{\psi_j(t)},
\end{equation}
where $\star$ represents the particular unraveling method considered. The corresponding jump probabilities $p^{\star}_{\psi_j(t)}$ depend on this choice and
are always evaluated for the target state $\ket{\psi_j(t)}$ of the reverse jump with $\ket{\psi_i(t)}$ being the destination of an earlier direct jump from $\ket{\psi_j(t)}$, in a given simulation scheme.
%depend on the particular unraveling scheme considered and are always evaluated from the target state $\ket{\psi_j(t)}$, with $\ket{\psi_i(t)}$ being the destination of a direct jump from $\ket{\psi_j(t)}$.

\subsection{$W$-rate operator}
\label{subsec:W}
Another possible approach to deal with non-Markovian dynamics consists of considering jump process different from the MCWF.
In particular, these require the definition of the so-called rate operator (RO) and are therefore known as rate operator quantum jumps (ROQJ) techniques.
A first method is the $W$-ROQJ, with the $W$-RO defined as \cite{Smirne2020, Diosi-orthogonal-jumps, Diosi-stochastic-repr}
\begin{equation}
    \label{eq:W}
    W_{\psi,t} \coloneqq (\id-\ketbra{\psi})\mathcal J_t\big[\ketbra\psi\big](\id-\ketbra{\psi}),
\end{equation}
where
\begin{equation}
    \label{eq:J_ME}
    \mathcal J_t[\rho]\coloneqq \sum_\alpha\gamma_\alpha(t) L_\alpha(t) \rho L_\alpha^\dagger(t)
\end{equation}
is the jump term of the master equation \eqref{eq:Lindblad}.
Suppose that the realization is in the state $\ket{\psi(t)}$, then the jump process reads \cite{Smirne2020}
\begin{equation}
    \label{eq:W_jump}
    \ket{\psi(t)}\mapsto\ket{\varphi^{\nu,W}_{\psi(t),t}},
\end{equation}
where $\ket{\varphi^{\nu,W}_{\psi(t),t}}$ is an eigenstate of $W_{\psi(t),t}$.
The jump probability is
\begin{equation}
    \label{eq:W_p_jump}
    p^{W}_{\nu} = \lambda^{\nu,W}_{\psi(t),t}\ dt
\end{equation}
where $\lambda^{\nu,W}_{\psi(t),t}$ is the corresponding eigenvalue, i.e.
\begin{equation}
    W_{\psi(t),t} = \sum_\nu\lambda^{\nu,W}_{\psi(t),t} \ketbra{\varphi^{\nu,W}_{\psi(t),t}}.
\end{equation}
Notice that such jumps are always to states orthogonal to the current state
$\braket{\psi(t)\vert\varphi^{\nu,W}_{\psi(t),t}} = 0$.
If no jumps happen, then the realization will evolve deterministically as
\begin{equation}
    \label{eq:W_det}
    \ket{\psi(t)}\mapsto\ket{\tilde\psi_{\text{det}}^W(t+dt)} = \left(\id - i K^W_{\psi(t),t}\ dt\right)\ket{\psi(t)},
\end{equation}
governed by the non-linear non-Hermitian effective Hamiltonian
\begin{equation}
    K^W_{\psi(t),t}\coloneqq K(t) + \frac i2\sum_\alpha\gamma_\alpha(t)\left(2 L_\alpha(t) \ell_{\alpha,\psi(t),t}^* - \abs{\ell_{\alpha,\psi(t),t}}^2\right),
\end{equation}
where $K(t)$ is the same effective Hamiltonian as in Eq.~\eqref{eq:MCWF_K}. The non-linear corrections read
\begin{equation}
    \ell_{\alpha,\psi(t),t}\coloneqq \braket{\psi(t)\vert L_\alpha(t)\vert \psi(t)}.
\end{equation}

The $W$-ROQJ unraveling technique can be applied for all dynamics which are P divisible, without requiring CP divisibility.
This follows from the fact that the P divisibility condition of Eq.~\eqref{eq:P_divisibility} is equivalent to the positivity of $W_{\psi,t}$ for all states $\psi$, and therefore the jump probabilities of Eq.~\eqref{eq:W_p_jump} are positive even for P but not CP divisible dynamics.
When, instead, the dynamics is non-P divisible, it is possible to equip the $W$-ROQJ unravelings with the reverse jumps of the NMQJ technique, where the source states of the jumps are now the eigenstates of $W_{\psi,t}$.

\subsection{$R$-rate operator}
\label{subsec:RO}
Another possible definition of the RO relies on the non-unique ways of writing the GKSL ME, both in the Markovian and in the non-Markovian regime.
The master equation \eqref{eq:Lindblad} can be written as
\begin{equation}
    \mathcal L_t[\rho] = \mathcal J_t[\rho] + \mathcal D_t[\rho],
\end{equation}
with $\mathcal J_t$ as in Eq.~\eqref{eq:J_ME} is the jump term, while
\begin{equation}
    \mathcal D_t[\rho]\coloneqq -i\left[K(t)\ \rho - \rho\ K^\dagger(t)\right]
\end{equation}
is the driving term, with $K(t)$ as in Eq.~\eqref{eq:MCWF_K}.
This subdivision is not unique, since the master equation is invariant under the transformations
\begin{gather}
    \label{eq:J_prime}
    \mathcal J_t[\rho]\mapsto \mathcal J^\prime_t[\rho]\coloneqq \mathcal J_t[\rho]+\frac12\left(C_t\ \rho + \rho\ C^\dagger_t\right),\\
    \label{eq:K_prime}
    K(t)\mapsto K^\prime(t)\coloneqq K(t)-\frac i2 C_t,
\end{gather}
where $C_t$ is an arbitrary time-dependent operator.

The corresponding unraveling scheme, denoted as $R$-ROQJ, consists of jumps
\begin{equation}
    \label{eq:R_jump}
    \ket{\psi(t)}\mapsto\ket{\varphi^{\nu,R}_{\psi(t),t}},
\end{equation}
to the eigenstates of the operator \cite{Chruscinski2022HowRepresentations}
\begin{equation}
    \label{eq:RO}
    \begin{split}
        R_{\psi(t),t}&\coloneqq \mathcal J^\prime_t[\ketbra{\psi(t)}]\\
        &= \mathcal J_t[\ketbra{\psi(t)}]+\frac12\left(C_t\ketbra{\psi(t)} + \ketbra{\psi(t)} C^\dagger_t\right)\\
        &=\sum_\nu\lambda^{\nu,R}_{\psi,t} \ketbra{\varphi^{\nu,R}_{\psi,t}}.
    \end{split}
\end{equation}
with probabilities given by the corresponding eigenvalue times the timestep $dt$
\begin{equation}
    \label{eq:R_p_jump}
    p^{R}_{\nu} = \lambda^{\nu,R}_{\psi(t),t}\ dt.
\end{equation}
The deterministic evolution reads
\begin{equation}
    \label{eq:R_det}
    \ket{\psi(t)}\mapsto\ket{\tilde\psi_{\text{det}}^R(t+dt)} = \left(\id - i K^R(t)\ dt\right)\ket{\psi(t)},
\end{equation}
with $K^R(t) = K^\prime(t)$ as in Eq.~\eqref{eq:K_prime}.
Notice that the operator $C_t$ defines a gauge freedom in the definition of the stochastic process and to each choice of $C_t$ correspond different stochastic realizations.
This freedom allows for an engineering of the stochastic processes, leading to effective ensembles with smaller number of states and therefore more efficient simulations.

The $R$-ROQJ technique is guaranteed to give positive jump probabilities, and therefore well-defined unravelings, whenever the dynamics is CP divisible for suitable choices of $C_t$.
If, instead, the dynamics is only P divisible, then $R_{\psi,t}$ has at most one negative eigenvalue for any transformation $C_t$ \cite{Chruscinski2022HowRepresentations}.

\subsection{Generalized rate operator}
\label{subsec:gen_RO}
The $R$-ROQJ method can be extended by noticing that, for any individual realization, it is possible to choose the transformation of Eqs.~\eqref{eq:J_prime}, \eqref{eq:K_prime} to depend not only on time but also on the state $\ket{\psi(t)}$ of the realization while preserving the generator $\mathcal L_t$.
Therefore, the RO of Eq.~\eqref{eq:RO} can be generalized by considering arbitrary transformations $C_{\psi(t),t}$ depending both on time and on the current state of the realization.
Upon defining $\ket{\phi_{\psi(t),t}} \coloneqq C_{\psi(t),t}\ket{\psi(t)}$, the generalized RO reads \cite{Settimo-RO, Settimo-SSE-RO_misc}
\begin{equation}
    \Psi\text{-}R_{\psi(t),t}\coloneqq \mathcal J_t[\ketbra{\psi(t)}]+\frac12\left(\ket{\phi_{\psi(t),t}}\bra{\psi(t)} + \ket{\psi(t)} \bra{\phi_{\psi(t),t}}\right).
\end{equation}
Notice that the vector $\ket{\phi_{\psi(t),t}}$ is not necessarily normalized and the RO is not invariant under a change in phase for $\ket{\phi_{\psi(t),t}}$.

The unravelings with the generalized RO ($\Psi$-ROQJ) are done in an analogous way to the $R$-ROQJ: the jumps happen to the eigenstates of $\Psi\text{-}R_{\psi(t),t}$
\begin{equation}
    \label{eq:Psi-R_jump}
    \ket{\psi(t)}\mapsto\ket{\varphi^{\nu,\Psi}_{\psi(t),t}}
\end{equation}
with probability given by the corresponding eigenvalue
\begin{equation}
    \label{eq:Psi-R_p_jump}
    p^{\Psi}_{\nu} = \lambda^{\nu,\Psi}_{\psi(t),t}\ dt.
\end{equation}
The deterministic evolution is driven by the non-Hermitian non-linear Hamiltonian
\begin{equation}
    \label{eq:K_Psi-R}
    K_{\psi(t),t}^\Psi\coloneqq K(t)-\frac i2\ket{\phi_{\psi(t),t}}\bra{\psi(t)}.
\end{equation}

The $\Psi$-ROQJ unraveling scheme extends the flexibility of the $R$-ROQJ in engineering the stochastic realizations, allowing for non-linear deterministic evolutions and smaller effective ensembles, thus giving more efficient simulations.
Furthermore, it encompasses both the $R$-ROQJ and the $W$-ROQJ as special cases.
The inclusion of the former is trivial by considering a transformation depending only on time, while the inclusion of the latter is obtained by fixing the transformation $\ket{\phi_{\psi(t),t}}$ such that $\ket{\psi(t)}$ is an eigenstate of the $\Psi$-RO with positive eigenvalue.
Thus, the $\Psi$-ROQJ technique gives positive rates for all P divisible dynamics, with the unravelings not requiring reverse jumps.
In \cite{Settimo-RO}, it was shown that there exist also some non-P divisible dynamics for which reverse jumps are not required, although this does not hold for all non-P divisible dynamics.

% --------------------------------------- Extension Hilbert space ---------------------------------------
\section{Extended Hilbert space}
\label{sec:extended_Hilbert}

Another possible way of applying stochastic processes with pure states realizations to non-Markovian dynamics is to consider extended Hilbert spaces.
Here we provide an overview of some widely used methods.

\subsection{Doubled Hilbert space}
\label{subsec:double_Hilbert}

One possible way to extend the Hilbert space to apply unravelings to arbitrary non-Markovian dynamics is to consider a doubled Hilbert space.
Such approach deals with master equations of the form \cite{Breuer1998Doubled, Breuer1999}
\begin{equation}
    \label{eq:ME_doubled}
    \frac d{dt}\rho = A(t)\rho + \rho B^\dagger(t) + \sum_i C_i(t)\rho D^\dagger_i(t),
\end{equation}
with $A(t)$, $B(t)$, $C_i(t)$, and $D_i(t)$ arbitrary linear and time-dependent operators.
Such equation reduces to the GKSL master equation \eqref{eq:Lindblad} for suitable choice of the operators.

The state of the system is described not by a single stochastic vector, but by a pair $\ket\theta = (\ket\phi,\ket\psi)^\top \in \mathscr H_2 =\mathscr H\oplus\mathscr H\cong \mathscr H\otimes\mathbb C^2$, where $\mathscr H$ is the system's Hilbert space.
The averaging process of Eq.~\eqref{eq:avg_solution} is modified accordingly, with the reduced dynamics obtained as
\begin{equation}
    \rho(t) = \int\mathcal D\theta\ p(\theta, t)\ \ket{\phi(t)}\bra{\psi(t)}.
\end{equation}
Upon defining
\begin{equation}
    F(t)=\begin{pmatrix}
        A(t) & 0\\
        0 & B(t)
    \end{pmatrix},\qquad
    J_i(t)=\begin{pmatrix}
        C_i(t) & 0\\
        0 & D_i(t)
    \end{pmatrix},
\end{equation}
the stochastic state vector on the extended Hilbert space evolves as
\begin{equation}
    \label{eq:doubled_SSE}
    d\ket\theta = -i G(\theta,t)\ dt+\sum_i \left(\frac{\norm{\ket{\theta}}}{\norm{J_i(t)\ket\theta}}J_i(t)\ket\theta - \ket\theta\right) dN_i^{\text{d}}(t),
\end{equation}
where $dN_i^{\text{d}}(t)$ are independent Poisson processes satisfying $dN_{i}^{\text{d}}(t)\, dN_{j}^{\text{d}}(t) = \delta_{i,j}\, dN_{i}^{\text{d}}(t)$ and
\begin{equation}
    \label{eq:doubled_Poisson}
    \mathbb E\left[dN_{i}^{\text{d}}(t)\right] = \frac{\norm{J_i(t)\ket\theta}^2}{\norm{\ket\theta}^2}dt,
\end{equation}
and
\begin{equation}
    G(\theta,t) = i\left(F(t) + \frac12\sum_i\frac{\norm{J_i(t)\ket\theta}^2}{\norm{\ket\theta}^2}\right)\ket\theta.
\end{equation}
Notice that the stochastic equation \eqref{eq:doubled_SSE} can be equivalently written as two independent equations, one for the ket $\ket\phi$ and one for the bra $\bra\psi$.

This unraveling scheme can be applied independently of the divisibility of the dynamical map, since the Poisson process of Eq.~\eqref{eq:doubled_Poisson} is always well defined and therefore the jump probabilities are always guaranteed to be positive.
Furthermore, upon suitable modification, this method can be applied to also compute time-ordered multitime correlation functions \cite{Breuer1998HeisenbergUnraveling} $\braket{\psi_0\vert O_n(t_n)\cdots O_1(t_1)\vert\psi_0}$, with $O_j$ arbitrary system operators and times $0\le t_1\le\ldots\le t_n$.

An alternative approach involving the same extension of the Hilbert space can be obtained, assuming $A(t) = B(t)$, by considering averaging of the form \cite{Kleinekathofer2002StochasticClass}
\begin{equation}
    \rho(t) = \int\mathcal D\theta\ p(\theta, t)\ \big[\ket{\phi(t)}\bra{\psi(t)} + \ket{\psi(t)}\bra{\phi(t)}\big].
\end{equation}
The stochastic vectors evolve as
\begin{gather}
    \begin{split}
        d\ket\psi =& \left(A(t) + \sum_i \frac{p^1_i(t)+p^2_i(t)}2\right)\ket\psi\ dt\\
        &+ \sum_i\left[\left(\frac{D_i(t)}{\sqrt{p^1_i(t)}}-1\right)dN^1_i(t) + \left(\frac{C_i(t)}{\sqrt{p^2_i(t)}}-1\right)dN^2_i(t)\right]\ket\psi,
    \end{split}
    \\
    \begin{split}
        d\ket\phi =& \left(A(t) + \sum_i \frac{p^1_i(t)+p^2_i(t)}2\right)\ket\phi\ dt\\
        &+ \sum_i\left[\left(\frac{C_i(t)}{\sqrt{p^1_i(t)}}-1\right)dN^1_i(t) + \left(\frac{D_i(t)}{\sqrt{p^2_i(t)}}-1\right)dN^2_i(t)\right]\ket\phi,
    \end{split}
\end{gather}
where $dN^k_i(t)$ are independent Poisson processes with $\mathbb E[dN^k_i(t)] = p^k_i(t)\, dt$ and the jump rates $p^{1,2}_i(t)$ are free parameters.
Such parameters can be optimized in order to improve the convergence of the methods, yielding optimal results when the norm of $\ket\psi\bra\phi+\ket\phi\bra\psi$ is preserved, i.e.
\begin{equation}
    \frac d{dt}\tr\big[\ket\psi\bra\phi+\ket\phi\bra\psi\big] = 0,
\end{equation}
which results in the additional constraints
\begin{equation}
    A(t) + A^\dagger(t) = -\sum_i\left(D_i^\dagger(t) C_i(t) + C_i^\dagger(t)D_i(t)\right)
\end{equation}
and fixes the free parameters as
\begin{gather}
    p^1_i(t) = \frac{\braket{\phi\vert C_i^\dagger(t) D_i(t)\vert\psi }+\braket{\psi\vert D_i^\dagger(t)C_i(t)\vert\phi}}{\braket{\phi\vert\psi} + \braket{\psi\vert\phi}},\\
    p^2_i(t) = \frac{\braket{\phi\vert D_i^\dagger(t) C_i(t)\vert\psi }+\braket{\psi\vert C_i^\dagger(t)D_i(t)\vert\phi}}{\braket{\phi\vert\psi} + \braket{\psi\vert\phi}}.
\end{gather}
This condition does not lead to positive jump rates $p_i^k(t)$. However, since they are arbitrary, they can be replaced by their absolute values.
The price to pay is that one has to introduce an additional weight factor for the trajectories which jumps between one and minus one \cite{Kleinekathofer2002StochasticClass}.

\subsection{Tripled Hilbert space}
\label{subsec:triple_Hilbert}
Another possible extension of the Hilbert space can be obtained by considering the tripled Hilbert space $\mathscr H_3 = \mathscr H\oplus \mathscr H\oplus \mathscr H \cong \mathscr H\otimes\mathbb C^3$.
Let us assume, for the sake of simplicity, that the reduced density matrix evolves via a master equation of the form \eqref{eq:ME_doubled} with only two jump operators $C(t)$ and $D(t)$
\begin{equation}
    \label{eq:ME_tripled}
    \begin{split}
        \frac d{dt}\rho =& -i[H(t),\rho] + C(t)\rho D^\dagger(t) + D(t)\rho C^\dagger(t)\\
        &- \frac 12\{D^\dagger(t)C(t)+C^\dagger(t) D(t),\rho\}.
    \end{split}
\end{equation}
In \cite{Breuer2004}, it was shown that it is possible to construct a density matrix $W(t)$ on the tripled Hilbert $\mathscr H_3$ space such that the solution of \eqref{eq:ME_tripled} can be obtained as
\begin{equation}
    \label{eq:rho_S_tripled}
    \rho(t) = \frac{{\braket{1\vert W(t)\vert2}}}{\tr{\braket{1\vert W(t)\vert2}}},
\end{equation}
where $\{\ket1,\ket2,\ket3\}$ is an orthonormal basis of $\mathbb C^3$, with the initial condition
\begin{equation}
    W(0) = \rho(0)\otimes\ketbra\chi,\qquad\ket\chi = \frac1{\sqrt2}(\ket1+\ket2).
\end{equation}
This density matrix $W(t)$ on the enlarged Hilbert space evolves according to the following Markovian master equation
\begin{equation}
    \frac{d}{dt}W = -i\left[\bar H^{(3)}(t),\rho\right] + \sum_{i=1}^4 J_i(t)\rho J_i^\dagger(t)-\frac12\{J_i^\dagger(t)J_i(t),\rho\},
\end{equation}
where
\begin{gather}
    \label{eq:J_1_tripled}
    J_1(t) = C(t)\otimes\ketbra1 + D(t)\otimes\ketbra2,\\
    J_2(t) = D(t)\otimes\ketbra1 + C(t)\otimes\ketbra2,\\
    J_3(t) = \Omega(t)\otimes\ket3\bra1,\\\label{eq:J_4_tripled}
    J_4(t) = \Omega(t)\otimes\ket3\bra2,\\
    \bar H^{(3)}(t) = H(t) \otimes \id,
\end{gather}
and the operator $\Omega(t)$ is such that
\begin{equation}
    \label{eq:Omega_tripled}
    \Omega^\dagger(t)\Omega(t) = a(t)\id - \left[C(t)-D(t)\right]^\dagger\left[C(t)-D(t)\right],
\end{equation}
with $a(t)$ being an arbitrary non-negative number such that
\begin{equation}
    \label{eq:a_tripled}
    a(t)\ge\norm{C(t)-D(t)}_\infty,
\end{equation}
where $\norm{\cdot}_\infty$ is the operator norm, i.e. the largest eigenvalue in absolute value.
Notice that Eq.~\eqref{eq:a_tripled} implies that a solution of Eq.~\eqref{eq:Omega_tripled} is guaranteed to exist.
However such solution is not unique, since for any unitary $U(t)$, $U(t)\Omega(t)$ is also a solution.
This construction can be generalized to arbitrary sets of operators $\{C_i(t)\}$ and $\{D_i(t)\}$ by considering a set of operators $\Omega_i(t)$ and $J_{\alpha,i}(t)$, $\alpha=1,\ldots,4$, defined analogously to Eqs.~\eqref{eq:J_1_tripled}-\eqref{eq:J_4_tripled}, \eqref{eq:Omega_tripled}.
Since the master equation \eqref{eq:ME_tripled} is Markovian, it can be unraveled with the MCWF and, unlike the doubled Hilbert space, the ket $\ket\psi$ and the bra $\bra\psi$ evolve in the same way.

Furthermore, this stochastic unraveling technique shows the existence of genuine quantum trajectories on $\mathscr H_3$, even for dynamics that are non-Markovian in $\mathscr H$.
Or, in other words, the stochastic process unraveling the master equation \eqref{eq:ME_tripled} can be interpreted as the outcome of a continuous measurement process.
However, this continuous measurement on $\mathscr H_3$ does not imply a continuous measurement on $\mathscr H$, since the process $W(t)\mapsto\rho(t)$ of Eq.~\eqref{eq:rho_S_tripled} is not linear.

\subsection{Influence martingale}
\label{subsec:influence_martingale}
Another possible approach is to extend the space to not only include the state vector but also a martingale $\mu(t)$, i.e. a a stochastic process such that the conditional expectation of the future value, given all past information, is equal to the current value \cite{Williams1991ProbabilityMartingales, Klebaner2012IntroductionApplications}.
Such martingale, known as the influence martingale (IM), enters in the stochastic process by changing the averaging of Eq.~\eqref{eq:avg_solution} as \cite{Donvil2022, Donvil2023Unraveling-pairedChannels}
\begin{equation}
    \label{eq:avg_influence_martingale}
    \rho(t) = \mathbb E\left[\mu(t)\, \ketbra{\psi(t)}\right].
\end{equation}
Notice that such averaging preserves the trace only on average.
The state vectors evolve as
\begin{equation}
    \label{eq:influence_martingale_state}
    \begin{split}
        d\ket{\psi(t)} = &-iK(t) \ket{\psi(t)}\, dt - \frac12\sum_\alpha\gamma_\alpha(t)\norm{L_\alpha(t)\ket{\psi(t)}}^2\ket{\psi(t)}\, dt\\
        &+\sum_\alpha\left(\frac{L_\alpha(t)\ket{\psi(t)}}{\norm{L_\alpha(t)\ket{\psi(t)}}}-\ket{\psi(t)}\right) dN_{\alpha}^{\text{IM}}(t),
    \end{split}
\end{equation}
where $dN_\alpha^{\text{IM}}(t)$ are independent Poisson processes satisfying $dN_{\alpha}^{\text{IM}}(t)\, dN_{\beta}^{\text{IM}}(t) = \delta_{\alpha,\beta}\, dN_{\alpha}^{\text{IM}}(t)$ and
\begin{equation}
    \mathbb E\left[dN_{\alpha}^{\text{IM}}(t)\big\vert\ket{\psi(t)}\right] = r_{\alpha}(t) \norm{L_\alpha(t)\ket{\psi(t)}}^2\ dt,
\end{equation}
with $r_{\alpha}(t)$ being strictly positive real functions.
The IM obeys the following stochastic equations
\begin{gather}
    \label{eq:influence_martingale_weight}
    d\mu(t) = \mu(t)\sum_\alpha \left(\frac{\gamma_\alpha(t)}{r_\alpha(t)}-1\right)d\iota_\alpha(t),\qquad \mu(0) = 1,\\
    d\iota_\alpha(t) = dN_\alpha(t) - r_{\alpha}(t) \norm{L_\alpha(t)\ket{\psi(t)}}^2\ dt.
\end{gather}
This method does not require any divisibility condition of the dynamics, since violations of divisibility can be encoded in non-positive values of the weight martingale $\mu(t)$.
Also, the extension needed to perform the simulation is minimal, since one only needs include in the description only one extra real parameter $\mu(t)$.

\subsection{Pseudo-Lindblad quantum trajectory}
\label{subsec:PLQJ}
Another approach requiring as extension only a single real number is the so-called pseudo-Lindblad quantum trajectory (PLQT) approach \cite{Becker2023}.
The state vector $\ket{\psi(t)}$ evolves as in the MCWF unravelings, but with jump probabilities
\begin{equation}
    \label{eq:PLQT_p_jump}
    p^{{\text{PL}}}_\alpha = \abs{\gamma_\alpha(t)} \norm{L_\alpha(t)\ket{\psi(t)}}^2\ dt
\end{equation}
which are always positive.
The Hilbert space is extended by a single bit $s(t)=\pm1$, such that $s(0) = 1$ and 
\begin{equation}
    s(t+dt) = \frac{\gamma_\alpha(t)}{\abs{\gamma_\alpha(t)}} s(t)
\end{equation}
if the jump $\ket{\psi(t)}\mapsto\ket{\psi_\alpha^{\text{MC}}(t+dt)}$ occurred, while $s(t+dt) = s(t)$ if the realization evolved deterministically.
Notice that $s(t)$ can change sign only if the rate is negative.

The reduced state is obtained as
\begin{equation}
    \rho(t) = \mathbb E\left[s(t)\, \ketbra{\psi(t)}\right]
\end{equation}
and, like for the IM, the trace is preserved only on average.
Because of this, the method exhibits an algorithmic relaxation time after which the simulation becomes unstable.
Such relaxation time corresponds to the time at which the number of realizations with $s(t) = +1$ roughly equals the number of realizations with $s(t) = -1$.

% --------------------------------------- Examples ---------------------------------------
\section{Examples}
\label{sec:examples}

In this section, we apply and compare the unraveling methods of Sec.~\ref{sec:Hilbert_space} and \ref{sec:extended_Hilbert} to some common dynamical maps.

\subsection{Qubit example for the different methods}
\label{subsec:example_qubit}
In order to exemplify the unraveling techniques, we consider a common dynamics to which such unraveling techniques are applied.
For the sake of example, we fix the dynamics to be of the phase covariant form \cite{Haase2018, Smirne2016UltimateEstimation}, for which the master equation reads
\begin{equation}
    \label{eq:ME_ph_cov}
    \mathcal L_t[\rho] = -i\,\omega_0\,[\sigma_z, \rho] + \sum_{\alpha=\pm,z}\gamma_\alpha(t) \sigma_\alpha \rho \sigma_\alpha^\dagger - \frac12\{\Gamma,\rho\},
\end{equation}
where $\gamma_\alpha(t)$ are time-dependent rates, $\sigma_+ = \ket1\bra0 = \sigma_-^\dagger$, and $\sigma_z = \ketbra1-\ketbra0$.
Such  master equation gives CP divisible dynamics if and only if $\gamma_\alpha(t)\ge0$, while P divisibility is equivalent to \cite{Filippov2020, Teittinen2018}
\begin{equation}
    \label{eq:ph_cov_P_div}
    \gamma_\pm(t)\ge0\qquad\text{and}\qquad\gamma_z(t)\ge-\frac12\sqrt{\gamma_+(t)\gamma_-(t)}.
\end{equation}
First, we consider a P but not CP divisible dynamics, for which all unraveling schemes can be applied.
Then, we move to a master equation for which P and CP divisibility are violated and we restrict to the methods which can deal with such class of dynamics.

\subsubsection{Eternally non-Markovian}
\label{subsubsec:ph_cov_P_div}
\begin{figure}
    \centering
    \includegraphics[width=\linewidth]{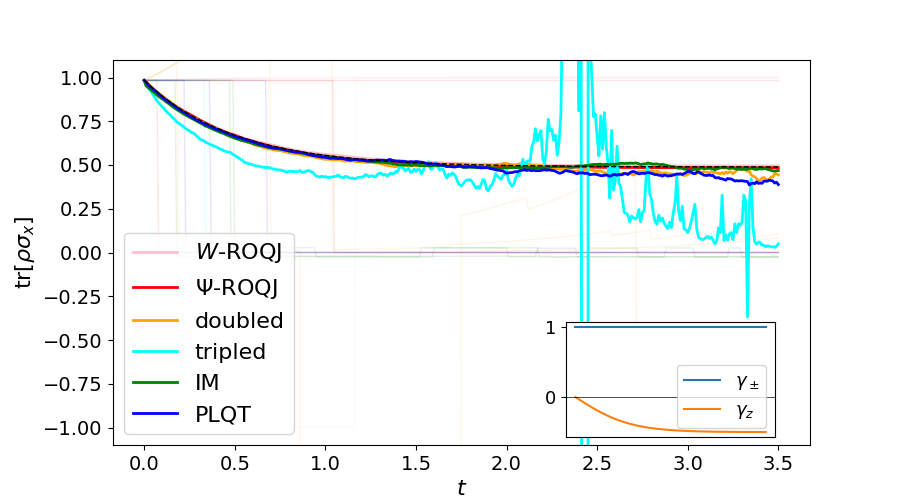}
    \caption{Eternally non-Markovian dynamics of Eq.~\eqref{eq:enm}.
    For each method, $10^4$ stochastic trajectories have been used and a timestep $dt=10^{-2}$.
    Three stochastic realizations are shown in lighter shade.
    The execution times are: $W$-ROQJ: 1447 ms, $\Psi$-ROQJ: 15 ms, doubled Hilbert space: 264 ms, tripled Hilbert space: 1809 ms, IM: 144 ms, PLQT: 59 ms.
    Inset: rates.}
    \label{fig:enm}
\end{figure}

As a first example of phase covariant dynamics to which apply the unravelings, we consider the so-called eternally non-Markovian dynamics \cite{Hall2014, Megier2017}, for which the rates read
\begin{equation}
    \label{eq:enm}
    \gamma_\pm(t) = 1,\qquad\gamma_z(t) = -\frac12\tanh t.
\end{equation}
This dynamics is P divisible, but CP divisibility is violated since $t=0$ because of the negativity of $\gamma_z(t)$.
In Fig.~\ref{fig:enm}, we show a comparison of unraveling techniques of Sec.~\ref{sec:Hilbert_space} and \ref{sec:extended_Hilbert}.
All methods were unraveled using the same numerical parameters and the code used is available at \cite{github}.
The NMQJ is not applicable on this example because of the need of reverse jumps since the initial time.

Notice that, for this example, the $R$-ROQJ and $\Psi$-ROQJ unravelings coincide, since we are using a state-independent transformation $C_t=\gamma_z(t)\id$ as in Eqs.~\eqref{eq:J_prime}, \eqref{eq:K_prime}.
This choice of the transformation allows for a small effective ensemble: only the deterministically evolving state and the states $\ket0$ and $\ket1$ are necessary to describe the exact solution, thus drastically enhancing the computational performance.

The tripled Hilbert space method is numerically unstable for this example, and the reason is the negativity of the rate since $t=0$.
Such instability, in fact, is less prominent if CP divisibility is violated only at later times.

The PLQT, instead, presents an algorithmic relaxation time after which the simulations become unstable, as noted in \cite{Becker2023}.
This relaxation time, however, is larger than the timescale after which the system reaches its steady state.

\subsubsection{Non-P-divisible dynamics}
\label{subsubsec:ph_cov_non_P_div}
\begin{figure}
    \centering
    \includegraphics[width=\linewidth]{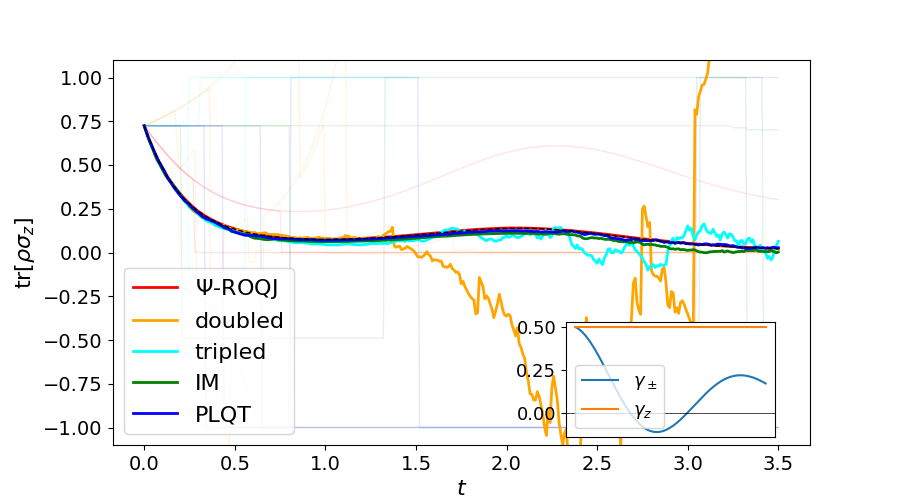}
    \caption{Non P divisible phase covariant dynamics of Eq.~\eqref{eq:non_P_div_rates} for $\kappa = 1/4$.
    For each method, $10^4$ stochastic trajectories have been used and a timestep $dt=10^{-2}$.
    Three stochastic realizations are shown in lighter shade.
    The execution times are: $\Psi$-ROQJ: 33 ms, doubled Hilbert space: 272 ms, tripled Hilbert space: 1816 ms, IM: 145 ms, PLQT: 61 ms.
    Inset: rates.}
    \label{fig:non-P-div}
\end{figure}

As a second example, we consider a dynamics which is not P-divisible, but nevertheless it is possible to apply the $\Psi$-ROQJ unravelings without reverse jumps.
Since P divisibility is violated, both the $W$-ROQJ and the $R$-ROQJ cannot be applied without using reverse jumps.
We choose the rates as \cite{Settimo-RO}
\begin{equation}
    \label{eq:non_P_div_rates}
    \gamma_\pm(t) = \frac12 e^{-t/10}\left[\kappa + (1-\kappa)e^{-t/4}\cos(2t)\right],\qquad\gamma_z(t) = \frac12
\end{equation}
and the unravelings are shown in Fig.~\ref{fig:non-P-div}.

Similarly to the previous example, the $\Psi$-ROQJ allows for a small effective ensemble consisting of the deterministically evolving state and the eigenstates of $\sigma_x$, thus drastically improving the simulation efficiency.

Both the doubled and tripled Hilbert space present numerical instability.
For the doubled, the instability begins when P divisibility is violated
This instability is due to the fact that the norm of $\phi$ and $\psi$ are not individually preserved.
For the tripled Hilbert space, the instability becomes noticeable only at later times.

The other methods give qualitatively similar results and stochastic realizations.

\subsection{Example of different physical systems}
\label{subsec:example_diff_sys}
In this section, we briefly survey works in which stochastic unravelings have been applied to realistic non-Markovian models, highlighting the physical motivation and the advantages of the methods.

The NMQJ has been widely employed to unravel non-Markovian open system dynamics.
However, for it to be efficient, a small effective ensemble is necessary.
Nevertheless, it was shown that it can deal with both two and three level systems \cite{Piilo2008, Piilo2009OpenJumps}, as well as two level driven systems \cite{Luoma-diffusive-NMQJ}.
This method was also employed to investigate excitonic energy transfer in photosynthetic complexes \cite{Rebentrost2009, Ai2013, Ai2014AnFields}, showing that non-Markovian effects lead to increased exciton transport.
Suitable generalizations to the method allowed for its application to many-body system \cite{Chiriaco2023}, by deriving the probability of the stochastic realizations via diagrammatic models having a structure similar to that of a Dyson equation.

The $W$-ROQJ technique was shown to be able to unravel arbitrary P divisible dynamics, independently of the underlying physical system.
Noticeable applications of such method include the study of a seven-site system
with both Hamiltonian and dissipative interaction between them \cite{Smirne2020}, as well as the eternally non-Markovian dynamics.
The $R$-ROQJ improved the simulation efficiency of the eternally non-Markovian dynamics by allowing a small effective ensemble for the description of the solution \cite{Chruscinski2022HowRepresentations}.

The results of both the $W$- and the $R$-ROQJ can be improved by the $\Psi$-ROQJ, which was shown to allow for an effective ensemble containing only three states for any qubit phase covariant dynamics, as well as the eternally non-Markovian with non-trivial driving breaking phase covariance \cite{Settimo-RO}.
It can also be used to unravel dynamics which violate P divisibility without requiring reverse jumps and also in cases in which it is violated since the beginning of the evolution \cite{Settimo-OPD}.

The doubled and tripled Hilbert space unravelings have been shown to be able to deal with master equations which are not necessarily in the GKSL form and were applied to simulate the spontaneous decay of two level systems \cite{Breuer1998Doubled}, the damped Jaynes-Cummings model \cite{Breuer2004}.
They can be also used to unravel multi-time correlations by applying the unravelings to matrix elements in the Heisenberg picture \cite{Breuer1998HeisenbergUnraveling}.
However, the numerical instability of the methods highlighted in Sec.~\ref{subsec:example_qubit} limited the application to more complex systems.

The IM method was originally shown to provide good description of the dynamics not only for qubit systems, but also for coupled large systems \cite{Donvil2022}.
The method can also be employed in error mitigation tasks on realistic quantum devices, in which the non-positive quasi-probability distribution is simulated $\mu(t)$ via suitable postprocessing \cite{Donvil2023QuantumMitigation_misc, Rossini2023SingleDynamics}.
More recent applications employed of suitable modifications of the IM formalism, or suitable modifications, to describe Non-Hermitian pseudomodes in strongly coupled systems \cite{Menczel2024Non-HermitianThermodynamics} and large spin chains \cite{Mondal2025_misc}.
Numerical implementations of the IM method are available on the QuTiP package \cite{qutip, qutip2}.

Lastly, the PLQT method was shown to be useful to unravel master equations both in GKSL and Redfield form, not only for qubits but also for the extended Hubbard chain of spinless fermions \cite{Becker2023}.
Nevertheless, this method might present numerical instability for non CP if the number of realizations with positive and negative weights become roughly equal.

% --------------------------------------- Beyond Lindblas ---------------------------------------
\section{Beyond GKSL}
\label{sec:beyond_Lindblad}

In this Section, we briefly described some ways in which stochastic unravelings can be applied in some situations in which the reduced dynamics is not described by master equations in the GKSL form of Eq.~\eqref{eq:Lindblad}.

\subsection{Initial correlations}
\label{subsec:init_corr}
One of the ways in which the reduced system dynamics can deviate from the GKSL description is if the global system and  environment state is initially entangled.
The assumption of initially uncorrelated system and environment, although simplifying the mathematical description, is not always physically realistic \cite{Pechukas1994, Alicki1995, Shaji2005, Carteret2008, Dominy2016} and many different strategies to include initial correlations in the reduced dynamics have been proposed \cite{Modi2012, Vacchini2016, Paz-Silva-B+, Colla2022}.

In particular, the reduced system dynamics does not need to be described by a CPTP map, and therefore stochastic unravelings cannot be applied in a straightforward way.
Nevertheless, the reduced dynamics can be described via a set of CPTP maps.
In order to derive such maps, a widely used method is the so-called bath positive (B+) or one-sided positive decomposition (OPD) \cite{Paz-Silva-B+}.
This relies on the fact that it is always possible to find a (over)complete system of system operators $\{Q_\alpha\}_\alpha$ such that
\begin{equation}
    \rho_{SE} = \sum_\alpha w_\alpha\, Q_\alpha\otimes\rho_\alpha,
\end{equation}
where $w_\alpha\ge0$ are non-negative real numbers and $\rho_\alpha$ are environmental operators.
The maximum number of terms present in the sum depends on the Schmidt rank of $\rho_{SE}$ and is always bound by $d^2$, where $d$ is the dimension of the system's Hilbert space \cite{Smirne2022b}.
For an explicit way of deriving the $Q_\alpha$ and the associated environmental states $\rho_\alpha$, see \cite{Paz-Silva-B+}.
The reduced dynamics will then be described by a sum of CPTP maps $\Phi_\alpha$, one for each $\rho_\alpha$, with weight $w_\alpha$, i.e.
\begin{equation}
    \rho(t) = \sum_\alpha w_\alpha\, \Phi^\alpha_t[Q_\alpha], \qquad \Phi^\alpha_t[\rho]\coloneqq \tr_E\left[U(t)\, \rho\otimes\rho_\alpha\, U^\dagger(t)\right],
\end{equation}
where $U(t)$ is the global unitary evolution.

From the CPTP maps $\Phi_\alpha$ it is possible to derive the Lindblad operators $\mathcal L^\alpha_t = \Phi_t^\alpha\circ{(\Phi_t^\alpha)}^{-1}$.
However, the initial conditions $Q_\alpha$ are in general not positive (positivity of the $Q_\alpha$ implies that the global state is separable \cite{Bengtsson2006}) and therefore they cannot be written as an average over positive operators as in Eq.~\eqref{eq:avg_solution}.
This fact would thus cause the failure of all the unraveling schemes.
A solution to this problem was derived in \cite{Settimo-OPD}, in which it was observed that it is always possible to write $Q_\alpha$ as a difference between states
\begin{equation}
    Q_\alpha = Q_\alpha^+-Q_\alpha^- = \mu_\alpha^+\Sigma_\alpha^+ - \mu_\alpha^-\Sigma_\alpha^-,
\end{equation}
where
\begin{equation}
    Q_\alpha^\pm = \frac12\left(\abs{Q_\alpha}\pm Q_\alpha\right)\ge0,\qquad \abs X = \sqrt{X^\dagger X}
\end{equation}
are the positive and negative part of $Q_\alpha$, and
\begin{equation}
    \mu_\alpha^\pm = \tr Q_\alpha^\pm\ge0,\qquad\Sigma_\alpha^\pm = \frac{Q_\alpha^\pm}{\mu_\alpha^\pm}.
\end{equation}
Therefore, the unravelings can be applied to the states $\Sigma_\alpha^\pm$, obtaining their time evolution $\Phi_\alpha^\pm[\Sigma_\alpha^\pm]$ which allows to obtain the reduced dynamics as
\begin{equation}
    \rho(t) = \sum_\alpha w_\alpha\left(\mu_\alpha^+\Phi^\alpha_t[\Sigma_\alpha^+]-\mu_\alpha^-\Phi^\alpha_t[\Sigma_\alpha^-]\right).
\end{equation}
Notice that the generators $\mathcal L^\alpha_t$ of the master equations for $\Sigma_\alpha^\pm$ can be obtained either exactly or via suitable generalizations of the projector operator technique \cite{Trevisan-APO}.

This method allows to obtain not only the reduced evolution starting from the entangled initial state $\rho_{SE}$, but also of all states 
\begin{equation}
    \rho^{\mathcal R}_{SE} = \frac{(\mathcal R\otimes\operatorname{id})\rho_{SE}}{\tr (\mathcal R\otimes\operatorname{id})\rho_{SE}}
\end{equation}
obtained by applying the CP local operation $\mathcal R$ on $\rho_{SE}$.
Their reduced evolution reads
\begin{equation}
    \rho^{\mathcal R}_S(t) = \frac{\sum_{\alpha,\alpha^\prime}w_\alpha R_{\alpha,\alpha^\prime}\left(\mu_{\alpha^\prime}^+\Phi^\alpha_t\left[\Sigma_{\alpha^\prime}^+\right] - \mu_{\alpha^\prime}^-\Phi^\alpha_t\left[\Sigma_{\alpha^\prime}^-\right]\right)}{\sum_\alpha w_\alpha\tr\mathcal R\left[Q_\alpha\right]},
\end{equation}
where $R_{\alpha,\alpha^\prime}$ is the expansion of $\mathcal R$ in the basis $\{Q_\alpha\}$.
The terms $\Phi^\alpha_t\left[\Sigma_{\alpha^\prime}^\pm\right]$ are estimated by unraveling the master equations $\mathcal L^\alpha_t$ with initial condition $\Sigma_{\alpha^\prime}^\pm$.

\subsection{Trace-non-preserving}
\label{subsec:TNP}

A second class of physically realistic master equations that are not in GKSL form are master equations that preserve positivity but not trace.
Such master equations are of the same form of Eq.~\eqref{eq:Lindblad}, but the operator $\Gamma(t)$ is an arbitrary self-adjoint operator, which does not need to be in the form of Eq.~\eqref{eq:Gamma}.
Master equations of this form are relevant, for instance, when describing the dynamics in the Heisenberg picture \cite{Alicki2007, Settimo-SchroHeis}, non-Hermitian Hamiltonians and exceptional points \cite{Minganti2019, Minganti2020, Gu2025_misc}, counting fields and counting statistics \cite{Bagrets2003, Flindt2005, Garrahan2010, Garrahan2011}.

These master equations can be unraveled by generalizing the cloning algorithm from the classical \cite{Giardina2006, Dean2009} to the quantum setting \cite{Carollo2020, Settimo2026}.
In order to take into account the variation of the trace, the number of stochastic realizations must vary over time, and the average of Eq.~\eqref{eq:avg_solution_approx} now reads
\begin{equation}
    \rho(t) = \frac1{N(0)}\sum_{i=1}^{N(t)}\ketbra{\psi_i(t)},
\end{equation}
where $N(t)$ is the number of trajectories at time $t$, with
\begin{equation}
    \tr\rho(t) = \frac{N(t)}{N(0)}\ne1.
\end{equation}

Here, we recall the cloning method applied to the MCWF unravelings of Sec.~\ref{subsubsec:MCWF}, but it can be generalized in a straightforward way to other unraveling schemes.
In this setting, the jump probabilities $p_\alpha^{\text{MC}}(t)$ and states $\ket{\tilde\psi_\alpha^{\text{MC}}(t+dt)}$ are the same as Eqs.~\eqref{eq:MCWF_jump}, \eqref{eq:MCWF_p_jump}.
The deterministically evolving state $\ket{\tilde\psi_{\text{det}}^{\text{MC}}(t+dt)}$ is also unchanged, however the probability of such evolution reads
\begin{equation}
    p_{\text{det}}^{\text{MC}}(t) = 1-\sum_\alpha p^{{\text{MC}}}_\alpha(t) - \Big\lvert{\braket{\psi(t)\vert\big(\Gamma_L(t)-\Gamma(t)\big)\vert\psi(t)}}\Big\rvert\,dt,
\end{equation}
where
\begin{equation}
    \Gamma_L(t)\coloneqq \sum_\alpha\gamma_\alpha(t)\,L_\alpha^\dagger(t) L_\alpha(t).
\end{equation}
Additionally, in order to take into account the non-constant number of trajectories, two more possible evolutions are added:
a new independent copy of the trajectory $\psi(t)$ can be created with probability
\begin{equation}
    p_{\text{c}}^{\text{MC}}(t) = \max\{0,\,\braket{\psi(t)\vert\big(\Gamma_L(t)-\Gamma(t)\big)\vert\psi(t)}\,dt\},
\end{equation}
or the trajectory can be destroyed with probability
\begin{equation}
    p_{\text{d}}^{\text{MC}}(t) = \max\{0,\,\braket{\psi(t)\vert\big(\Gamma(t)-\Gamma_L(t)\big)\vert\psi(t)}\,dt\}.
\end{equation}
This way, stochastic unravelings can be generalized to trace-non-preserving master equations by allowing for a non-constant number of stochastic realizations.
Notice that the numerical requirements can be kept unchanged by resampling the trajectories in order to have an effective ensemble of constant size, and just keeping track of the change in number of trajectory to vary the trace \cite{Carollo2020}.

\section{On the existence of a continuous measurement scheme}
\label{sec:measurement}
The most widely investigated question behind stochastic trajectories is: {\it given a stochastic realization, when can it be reproduced via a continuous measurement acting on the system?}
Whenever the dynamics is P or CP divisible, then such stochastic measurement scheme is always possible, since both the MCWF and the $W$-ROQJ realizations do indeed arise from a continuous monitoring.
The answer becomes more complicated for non P divisible dynamics.
Early results \cite{Diosi2008Non-markovianObservables} hinted at the existence of a continuous measurement for arbitrary non-Markovian dynamics, however the proposed measurement scheme was flawed in the sense that it required entangled measurement devices both in space and in time.
However, since the detection event disentangles the measurement apparata, even if the state at any given time is indeed a conditioned state, joining up these states as a trajectory creates a fiction \cite{Wiseman2008Pure-stateExist}.

Given an arbitrary non-Markovian dynamics, there always exists an unraveling, either diffusive \cite{Diosi-NMQSD, Gambetta2003} or jump like \cite{Gambetta-jump-nM, Luoma2011ConnectingSystems}.
In other words, there always exists a real stochastic processes $\{z_k(t)\}$ such that the reduced state can be written as an expectation value over such process
\begin{equation}
    \rho(t) = \mathbb E\left[\ketbra{\psi_{\{z_k(t)\}}}\right].
\end{equation}
The resulting dynamics fails to be a state conditioned over a continuous measurement whenever the dynamics is non-Markovian \cite{Gambetta2002Non-MarkovianTheory}, i.e. a genuine quantum trajectory.

Nevertheless, genuine quantum trajectories have been shown to exist in the tripled Hilbert space framework.
However, the trajectories on the enlarged space do not provide trajectories also on the original Hilbert space, since the process of Eq.~\eqref{eq:rho_S_tripled} for obtaining the reduced state is not described by a CP map.

On the other hand, P divisibility is not a necessary condition for the existence of a continuous measurement scheme.
Indeed, in \cite{Settimo-RO}, a dynamics which is not P divisible but allowing for a continuous measurement scheme was presented.
The continuous measurement used corresponded to the $\Psi$-ROQJ method.

Currently, the condition for the existence of a continuous measurement scheme is unknown.

\begin{table}[]
    \centering\footnotesize
    \begin{tabular}{c|c|c|c|c|c|c}
        Method & CP & P, non-CP & non-P & Indep. & Extension & Meas.\\ \hline
        MCWF & $\checkmark$ & $\times$ & $\times$ & $\checkmark$ & $\times$ & $\checkmark$ \\
        WTD & $\checkmark$ & $\times$ & $\times$ & $\checkmark$ & $\times$ & $\checkmark$\\
        NMQJ & $\checkmark$ & $\checkmark$ & $\checkmark$ & $\times$ & $\times$ & $\times$\\
        $W$-ROQJ & $\checkmark$ & $\checkmark$ & $\times$ & $\checkmark$ & $\times$ & $\checkmark$\\
        $R$-ROQJ & $\checkmark$ & ? & $\times$ & $\checkmark$ & $\times$ & $\checkmark$\\
        $\Psi$-ROQJ & $\checkmark$ & $\checkmark$ & ? & $\checkmark$ & $\times$ & $\checkmark$\\
        Doubled & $\checkmark$ & $\checkmark$ & $\checkmark$ & $\checkmark$ & $d$ & $\times$\\
        Tripled & $\checkmark$ & $\checkmark$ & $\checkmark$ & $\checkmark$ & $2\ d$ & $\checkmark$\\
        IM & $\checkmark$ & $\checkmark$ & $\checkmark$ & $\checkmark$ & $1$ & $\times$\\
        PLQT & $\checkmark$ & $\checkmark$ & $\checkmark$ & $\checkmark$ & $1$ & $\times$\\
    \end{tabular}
    \caption{Schematic comparison of the unraveling methods presented.
    The symbols read ``$\checkmark$": yes; ``$\times$": no; ``?": unknown.
    The first three columns (``CP"; P, ``non-CP", and ``non-P") describe the applicability of the unraveling method depending on the divisibility properties of the dynamical map.
    ``Indep." refers to whether the stochastic realizations are independent from each other.
    ``Extension" is the size of the (eventual) extension of the system's Hilbert space.
    Lastly, ``Meas." describes if the unraveling scheme corresponds to a continuous measurement applied to the system.}
    \label{tab:table_unr}
\end{table}

% --------------------------------------- Discussion ---------------------------------------
\section{Discussion and outlook}
\label{sec:discussion}
In recent years, significant progress has been made in simulating realistic open system dynamics beyond the memoryless assumption.
Among the various simulation techniques, stochastic unravelings play a crucial twofold role: on the one hand, they allow for powerful simulation techniques by reducing the dimensionality of the problem from $d\times d$ density matrices to $d$ dimensional stochastic state vectors.
On the other hand, they provide an alternative way of representing the reduced dynamics via pure state realizations obtained via continuous measurement processes.
In this work, we have presented different unraveling schemes for non-Markovian dynamics, with particular emphasis on how they allow for efficient simulations of open system dynamics.
A schematic comparison of the unraveling methods discussed in the present review is presented in Table \ref{tab:table_unr}.

However, the computational efficiency is not the only reason for the importance of stochastic unravelings.
Among the many applications, they can be used to detect non-Markovianity \cite{Luoma2014DetectingMonitoring}, to provide an operational interpretation of quantum memory as the existence of a series of conditional maps which unravel the dynamics \cite{Backer2024}, in the study of non-Hermitian pseudomodes \cite{Menczel2024Non-HermitianThermodynamics}, or in telling apart different unraveling given the continuous measurement record \cite{Pinol2024TellingAverages, Brown2025GaugeTrajectories, Gaona-Reyes2025TheoreticalUnravelings}.
They can also be extended to the Keldysh-contour formalism \cite{Cavina2025}.
Furthermore, their interest is not only theoretical, but they can also be reproduced experimentally \cite{Gleyzes2007, Minev2019, Basche1995, Peil1999, Vijay2011, Jelezko2002}.

Just like in the classical case, stochastic trajectories can be also studied from a thermodynamical point of view.
In order to do so, one has to introduce the notion of a reverse process.
By doing so, it is possible to assign thermodynamical quantities to such stochastic trajectories, like entropy production \cite{Garrahan2010, Leggio2013EntropyTrajectories, Perfetto2022}.

%For open system dynamics for which the Markov approximation holds, the time evolution can be efficiently simulated by means of the MCWF technique.
%However, in the realistic case where the approximation fails, the rates in the GKSL master equation would become temporarily negate and the MCWF technique would give rise to unphysical negative jump probabilities.
%Nevertheless, numerous unraveling techniques for the non-Markovian case have been developed in recent years, which allow one to overcome the limitation of negative rates.

In this review, we have presented and compared a broad class of quantum jump unravelings for non-Markovian open system dynamics.
A first class of unravelings modifies the stochastic process directly on the system's Hilbert space, without requiring any extra degrees of freedom.
The NMQJ technique overcomes the problem of the temporarily negative rates by introducing so-called reverse jumps, at the price of introducing correlations among different realizations.
The ROQJ family instead restores positivity by redefining the jump structure itself, preserving independence of realizations and, in suitable regimes, a continuous measurement interpretation.
In particular, the W- and R-ROQJ allow one to go beyond CP divisibility, while the generalized $\Psi$-ROQJ extends this feature even to certain non-P-divisible dynamics.
The possibility of engineering the rate operator enables the design of effective ensembles with few elements, leading to substantial numerical advantages.

A second class of methods circumvents the negativity of the rates by enlarging the Hilbert space.
The doubled and tripled Hilbert space techniques allow for positive jump rates on a space consisting of multiple copies of the system Hilbert space and the tripled case also allows for an interpretation as continuous measurement on the enlarged space.
The IM and PLQT techniques, instead, enable positive rates via a minimal extension consisting of just a single degree of freedom, either continuous (IM) or discrete (PLQT), which encodes the non-Markovian features of the dynamics.
Although these constructions require auxiliary degrees of freedom, the stochastic description still scales more favorably than direct density matrix simulation and remains efficient for moderate system sizes.

In summary, non-Markovian jump unravelings do not constitute a single method but rather a variety 
of techniques and approaches.
Different master equations, physical regimes and computational constraints favor different strategies.
Our unified comparison highlights both the strengths and the limitations of the available techniques, and we hope that it will provide a useful guide for selecting and further developing stochastic methods for realistic open quantum systems.

\section*{Acknowledgments}
FS acknowledges support from Magnus Ehrnroothin S\"a\"ati\"o.

\printbibliography

@article{Albarelli2024,
    title = {{A pedagogical introduction to continuously monitored quantum systems and measurement-based feedback}},
    year = {2024},
    journal = {Physics Letters, Section A: General, Atomic and Solid State Physics},
    author = {Albarelli, Francesco and Genoni, Marco G.},
    number = {December 2023},
    pages = {129260},
    volume = {494},
    publisher = {Elsevier B.V.},
    url = {https://doi.org/10.1016/j.physleta.2023.129260},
    doi = {10.1016/j.physleta.2023.129260},
    issn = {03759601},
    arxivId = {2312.13214}
}

@article{Trevisan-APO,
    title = {{Adapted projection operator technique for the treatment of initial correlations}},
    year = {2021},
    journal = {Physical Review A},
    author = {Trevisan, Andrea and Smirne, Andrea and Megier, Nina and Vacchini, Bassano},
    number = {5},
    volume = {104},
    publisher = {American Physical Society},
    doi = {10.1103/PhysRevA.104.052215},
    issn = {24699934},
    arxivId = {2107.13577}
}

@article{Kiilerich2016,
    title = {{Bayesian parameter estimation by continuous homodyne detection}},
    year = {2016},
    journal = {Physical Review A},
    author = {Kiilerich, Alexander Holm and M{\o}lmer, Klaus},
    number = {3},
    pages = {1--9},
    volume = {94},
    doi = {10.1103/PhysRevA.94.032103},
    issn = {24699934},
    arxivId = {1605.00902}
}

@article{Gammelmark2013,
    title = {{Bayesian parameter inference from continuously monitored quantum systems}},
    year = {2013},
    journal = {Physical Review A - Atomic, Molecular, and Optical Physics},
    author = {Gammelmark, Søren and M{\o}lmer, Klaus},
    number = {3},
    pages = {1--9},
    volume = {87},
    doi = {10.1103/PhysRevA.87.032115},
    issn = {10502947},
    arxivId = {1212.5700}
}

@article{Dominy2016,
    title = {{Beyond complete positivity}},
    year = {2016},
    journal = {Quantum Information Processing},
    author = {Dominy, Jason M. and Lidar, Daniel A.},
    number = {4},
    pages = {1349--1360},
    volume = {15},
    publisher = {Springer US},
    doi = {10.1007/s11128-015-1228-1},
    issn = {15700755},
    arxivId = {1503.05342}
}

@article{Hall2014,
    title = {{Canonical form of master equations and characterization of non-Markovianity}},
    year = {2014},
    journal = {Physical Review A - Atomic, Molecular, and Optical Physics},
    author = {Hall, Michael J.W. and Cresser, James D. and Li, Li and Andersson, Erika},
    number = {4},
    pages = {1--11},
    volume = {89},
    doi = {10.1103/PhysRevA.89.042120},
    issn = {10941622},
    arxivId = {1009.0845}
}

@article{Smith2002,
    title = {{Capture and release of a conditional state of a cavity QED system by quantum feedback}},
    year = {2002},
    journal = {Physical Review Letters},
    author = {Smith, W. P. and Reiner, J. E. and Orozco, L. A. and Kuhr, S. and Wiseman, H. M.},
    number = {13},
    pages = {1336011--1336014},
    volume = {89},
    doi = {10.1103/PhysRevLett.89.133601},
    issn = {00319007}
}

@article{Ai2013,
    title = {{Clustered Geometries Exploiting Quantum Coherence Effects for Efficient Energy Transfer in Light Harvesting}},
    year = {2013},
    journal = {The Journal of Physical Chemistry Letters},
    author = {Ai, Qing and Yen, Tzu-chi and Jin, Bih-yaw and Cheng, Yuan-chung},
    number = {15},
    month = {8},
    pages = {2577--2584},
    volume = {4},
    url = {https://pubs.acs.org/doi/10.1021/jz4011477},
    doi = {10.1021/jz4011477},
    issn = {1948-7185}
}

@article{BLPV-colloquim,
    title = {{Colloquium : Non-Markovian dynamics in open quantum systems}},
    year = {2016},
    journal = {Reviews of Modern Physics},
    author = {Breuer, Heinz-Peter and Laine, Elsi-Mari and Piilo, Jyrki and Vacchini, Bassano},
    number = {2},
    month = {4},
    pages = {021002},
    volume = {88},
    url = {https://link.aps.org/doi/10.1103/RevModPhys.88.021002},
    doi = {10.1103/RevModPhys.88.021002},
    issn = {0034-6861}
}

@article{Alicki1995,
    title = {{Comment on "reduced dynamics need not be completely positive"}},
    year = {1995},
    journal = {Physical Review Letters},
    author = {Alicki, Robert},
    number = {16},
    pages = {3020},
    volume = {75},
    doi = {10.1103/PhysRevLett.75.3020},
    issn = {00319007}
}

@article{Gorini1976,
    title = {{Completely positive dynamical semigroups of N -level systems}},
    year = {1976},
    journal = {Journal of Mathematical Physics},
    author = {Gorini, Vittorio and Kossakowski, Andrzej and Sudarshan, E. C. G.},
    number = {5},
    month = {5},
    pages = {821--825},
    volume = {17},
    url = {https://pubs.aip.org/jmp/article/17/5/821/225427/Completely-positive-dynamical-semigroups-of-N},
    doi = {10.1063/1.522979},
    issn = {0022-2488}
}

@article{Guevara2020,
    title = {{Completely positive quantum trajectories with applications to quantum state smoothing}},
    year = {2020},
    journal = {Physical Review A},
    author = {Guevara, Ivonne and Wiseman, Howard M.},
    number = {5},
    month = {11},
    pages = {052217},
    volume = {102},
    url = {https://link.aps.org/doi/10.1103/PhysRevA.102.052217},
    doi = {10.1103/PhysRevA.102.052217},
    issn = {2469-9926}
}

@article{White2020,
    title = {{Demonstration of non-Markovian process characterisation and control on a quantum processor}},
    year = {2020},
    journal = {Nature Communications},
    author = {White, G. A. L. and Hill, C. D. and Pollock, F. A. and Hollenberg, L. C. L. and Modi, K.},
    number = {1},
    month = {12},
    pages = {6301},
    volume = {11},
    publisher = {Springer US},
    url = {http://dx.doi.org/10.1038/s41467-020-20113-3 https://www.nature.com/articles/s41467-020-20113-3},
    doi = {10.1038/s41467-020-20113-3},
    issn = {2041-1723},
    pmid = {33298929},
    arxivId = {2004.14018}
}

@article{Riste2013,
    title = {{Deterministic entanglement of superconducting qubits by parity measurement and feedback}},
    year = {2013},
    journal = {Nature},
    author = {Rist{\`{e}}, D. and Dukalski, M. and Watson, C. A. and De Lange, G. and Tiggelman, M. J. and Blanter, Ya M. and Lehnert, K. W. and Schouten, R. N. and Dicarlo, L.},
    number = {7471},
    pages = {350--354},
    volume = {502},
    doi = {10.1038/nature12513},
    issn = {00280836},
    arxivId = {1306.4002}
}

@article{Chiriaco2023,
    title = {{Diagrammatic method for many-body non-Markovian dynamics: Memory effects and entanglement transitions}},
    year = {2023},
    journal = {Physical Review B},
    author = {Chiriac{\`{o}}, Giuliano and Tsitsishvili, Mikheil and Poletti, Dario and Fazio, Rosario and Dalmonte, Marcello},
    number = {7},
    month = {8},
    pages = {075151},
    volume = {108},
    url = {http://arxiv.org/abs/2302.10563 http://dx.doi.org/10.1103/PhysRevB.108.075151 https://link.aps.org/doi/10.1103/PhysRevB.108.075151},
    doi = {10.1103/PhysRevB.108.075151},
    issn = {2469-9950},
    arxivId = {2302.10563}
}

@article{Luoma-diffusive-NMQJ,
    title = {{Diffusive Limit of Non-Markovian Quantum Jumps}},
    year = {2020},
    journal = {Physical Review Letters},
    author = {Luoma, Kimmo and Strunz, Walter T. and Piilo, Jyrki},
    number = {15},
    pages = {150403},
    volume = {125},
    publisher = {American Physical Society},
    url = {https://doi.org/10.1103/PhysRevLett.125.150403},
    doi = {10.1103/PhysRevLett.125.150403},
    issn = {10797114},
    pmid = {33095626},
    arxivId = {2004.12072}
}

@article{Giardina2006,
    title = {{Direct evaluation of large-deviation functions}},
    year = {2006},
    journal = {Physical Review Letters},
    author = {Giardin{\`{a}}, Cristian and Kurchan, Jorge and Peliti, Luca},
    number = {12},
    pages = {1--4},
    volume = {96},
    doi = {10.1103/PhysRevLett.96.120603},
    issn = {00319007},
    arxivId = {cond-mat/0511248}
}

@article{Basche1995,
    title = {{Direct spectroscopic observation of quantum jumps of a single molecule}},
    year = {1995},
    journal = {Nature},
    author = {Basch{\'{e}}, Th. and Kummer, S. and Br{\"{a}}uchle, C.},
    number = {6510},
    month = {1},
    pages = {132--134},
    volume = {373},
    url = {https://www.nature.com/articles/373132a0},
    doi = {10.1038/373132a0},
    issn = {0028-0836}
}

@article{Settimo-SchroHeis,
    title = {{Divisibility of Dynamical Maps: Schr{\"{o}}dinger Versus Heisenberg Picture}},
    year = {2026},
    journal = {PRX Quantum},
    author = {Settimo, Federico and Smirne, Andrea and Luoma, Kimmo and Vacchini, Bassano and Piilo, Jyrki and Chru{\'{s}}ci{\'{n}}ski, Dariusz},
    number = {1},
    month = {2},
    pages = {010340},
    volume = {7},
    url = {https://link.aps.org/doi/10.1103/6dt2-sq44},
    doi = {10.1103/6dt2-sq44},
    issn = {2691-3399},
    arxivId = {2506.08103}
}

@article{Chruscinski2022,
    title = {{Dynamical maps beyond Markovian regime}},
    year = {2022},
    journal = {Physics Reports},
    author = {Chru{\'{s}}ci{\'{n}}ski, Dariusz},
    month = {12},
    pages = {1--85},
    volume = {992},
    url = {https://linkinghub.elsevier.com/retrieve/pii/S0370157322003428},
    doi = {10.1016/j.physrep.2022.09.003},
    issn = {03701573}
}

@article{Carteret2008,
    title = {{Dynamics beyond completely positive maps: Some properties and applications}},
    year = {2008},
    journal = {Physical Review A - Atomic, Molecular, and Optical Physics},
    author = {Carteret, Hilary A. and Terno, Daniel R. and Zyczkowski, Karol},
    number = {4},
    pages = {1--8},
    volume = {77},
    doi = {10.1103/PhysRevA.77.042113},
    issn = {10502947}
}

@article{Ficheux2018,
    title = {{Dynamics of a qubit while simultaneously monitoring its relaxation and dephasing}},
    year = {2018},
    journal = {Nature Communications},
    author = {Ficheux, Q. and Jezouin, S. and Leghtas, Z. and Huard, B.},
    number = {1},
    month = {5},
    pages = {1926},
    volume = {9},
    publisher = {Springer US},
    url = {http://dx.doi.org/10.1038/s41467-018-04372-9 https://www.nature.com/articles/s41467-018-04372-9},
    doi = {10.1038/s41467-018-04372-9},
    issn = {2041-1723},
    pmid = {29765040}
}

@article{Paz-Silva-B+,
    title = {{Dynamics of initially correlated open quantum systems: Theory and applications}},
    year = {2019},
    journal = {Physical Review A},
    author = {Paz-Silva, Gerardo A. and Hall, Michael J.W. and Wiseman, Howard M},
    number = {4},
    pages = {1--19},
    volume = {100},
    publisher = {American Physical Society},
    doi = {10.1103/PhysRevA.100.042120},
    issn = {24699934},
    arxivId = {1810.12540}
}

@article{DeVega2017,
    title = {{Dynamics of non-Markovian open quantum systems}},
    year = {2017},
    journal = {Reviews of Modern Physics},
    author = {De Vega, Inés and Alonso, Daniel},
    number = {1},
    pages = {1--58},
    volume = {89},
    doi = {10.1103/RevModPhys.89.015001},
    issn = {15390756},
    arxivId = {1511.06994}
}

@article{Settimo-OPD,
    title = {{Dynamics of open quantum systems with initial system-environment correlations via stochastic unravelings}},
    year = {2025},
    journal = {Physical Review A},
    author = {Settimo, Federico and Luoma, Kimmo and Chru{\'{s}}ci{\'{n}}ski, Dariusz and Smirne, Andrea and Vacchini, Bassano and Piilo, Jyrki},
    number = {4},
    month = {10},
    pages = {042204},
    volume = {112},
    url = {http://arxiv.org/abs/2502.12818 http://dx.doi.org/10.1103/q353-4232 https://link.aps.org/doi/10.1103/q353-4232},
    doi = {10.1103/q353-4232},
    issn = {2469-9926},
    arxivId = {2502.12818}
}

@article{RHP,
    title = {{Entanglement and Non-Markovianity of Quantum Evolutions}},
    year = {2010},
    journal = {Physical Review Letters},
    author = {Rivas, Ángel and Huelga, Susana F. and Plenio, Martin B.},
    number = {5},
    month = {7},
    pages = {050403},
    volume = {105},
    url = {https://link.aps.org/doi/10.1103/PhysRevLett.105.050403},
    doi = {10.1103/PhysRevLett.105.050403},
    issn = {0031-9007},
    arxivId = {0911.4270}
}

@article{Carollo2020,
    title = {{Entanglement statistics in Markovian open quantum systems: A matter of mutation and selection}},
    year = {2020},
    journal = {Physical Review E},
    author = {Carollo, Federico and P{\'{e}}rez-Espigares, Carlos},
    number = {3},
    pages = {1--6},
    volume = {102},
    publisher = {American Physical Society},
    doi = {10.1103/PhysRevE.102.030104},
    issn = {24700053},
    pmid = {33076006},
    arxivId = {1910.13940}
}

@article{Settimo-JSD,
    title = {{Entropic and trace-distance-based measures of non-Markovianity}},
    year = {2022},
    journal = {Physical Review A},
    author = {Settimo, Federico and Breuer, Heinz-Peter and Vacchini, Bassano},
    number = {4},
    month = {10},
    pages = {042212},
    volume = {106},
    publisher = {American Physical Society},
    url = {http://arxiv.org/abs/2207.13183 https://link.aps.org/doi/10.1103/PhysRevA.106.042212},
    doi = {10.1103/PhysRevA.106.042212},
    issn = {2469-9926},
    arxivId = {2207.13183}
}

@article{Buscemi2016,
    title = {{Equivalence between divisibility and monotonic decrease of information in classical and quantum stochastic processes}},
    year = {2016},
    journal = {Physical Review A},
    author = {Buscemi, Francesco and Datta, Nilanjana},
    number = {1},
    pages = {1--8},
    volume = {93},
    doi = {10.1103/PhysRevA.93.012101},
    issn = {24699934},
    arxivId = {1408.7062}
}

@article{Megier2017,
    title = {{Eternal non-Markovianity: from random unitary to Markov chain realisations}},
    year = {2017},
    journal = {Scientific Reports},
    author = {Megier, Nina and Chru{\'{s}}ci{\'{n}}ski, Dariusz and Piilo, Jyrki and Strunz, Walter T.},
    number = {1},
    month = {7},
    pages = {6379},
    volume = {7},
    url = {https://www.nature.com/articles/s41598-017-06059-5},
    isbn = {4159801706},
    doi = {10.1038/s41598-017-06059-5},
    issn = {2045-2322},
    pmid = {28743895},
    arxivId = {1608.07125}
}

@article{Goswami2021,
    title = {{Experimental characterization of a non-Markovian quantum process}},
    year = {2021},
    journal = {Physical Review A},
    author = {Goswami, K. and Giarmatzi, C. and Monterola, C. and Shrapnel, S. and Romero, J. and Costa, F.},
    number = {2},
    pages = {1--7},
    volume = {104},
    doi = {10.1103/PhysRevA.104.022432},
    issn = {24699934},
    arxivId = {2102.01327}
}

@article{Liu2011,
    title = {{Experimental control of the transition from Markovian to non-Markovian dynamics of open quantum systems}},
    year = {2011},
    journal = {Nature Physics},
    author = {Liu, Bi Heng and Li, Li and Huang, Yun Feng and Li, Chuan Feng and Guo, Guang Can and Laine, Elsi Mari and Breuer, Heinz Peter and Piilo, Jyrki},
    number = {12},
    pages = {931--934},
    volume = {7},
    doi = {10.1038/NPHYS2085},
    issn = {17452481},
    arxivId = {1109.2677}
}

@article{Livingston2022,
    title = {{Experimental demonstration of continuous quantum error correction}},
    year = {2022},
    journal = {Nature Communications},
    author = {Livingston, William P. and Blok, Machiel S. and Flurin, Emmanuel and Dressel, Justin and Jordan, Andrew N. and Siddiqi, Irfan},
    number = {1},
    pages = {1--7},
    volume = {13},
    publisher = {Springer US},
    doi = {10.1038/s41467-022-29906-0},
    issn = {20411723},
    pmid = {35484135},
    arxivId = {2107.11398}
}

@article{Liu2018,
    title = {{Experimental implementation of fully controlled dephasing dynamics and synthetic spectral densities}},
    year = {2018},
    journal = {Nature Communications},
    author = {Liu, Zhao-Di and Lyyra, Henri and Sun, Yong-Nan and Liu, Bi-Heng and Li, Chuan-Feng and Guo, Guang-Can and Maniscalco, Sabrina and Piilo, Jyrki},
    number = {1},
    month = {8},
    pages = {3453},
    volume = {9},
    publisher = {Springer US},
    url = {http://dx.doi.org/10.1038/s41467-018-05817-x https://www.nature.com/articles/s41467-018-05817-x},
    isbn = {4146701805},
    doi = {10.1038/s41467-018-05817-x},
    issn = {2041-1723},
    pmid = {30150668},
    arxivId = {1712.08071}
}

@article{Cialdi2019,
    title = {{Experimental investigation of the effect of classical noise on quantum non-Markovian dynamics}},
    year = {2019},
    journal = {Physical Review A},
    author = {Cialdi, Simone and Benedetti, Claudia and Tamascelli, Dario and Olivares, Stefano and Paris, Matteo G.A. and Vacchini, Bassano},
    number = {5},
    volume = {100},
    doi = {10.1103/PhysRevA.100.052104},
    issn = {24699934},
    arxivId = {1909.01113}
}

@article{Rybarczyk2015,
    title = {{Forward-backward analysis of the photon-number evolution in a cavity}},
    year = {2015},
    journal = {Physical Review A - Atomic, Molecular, and Optical Physics},
    author = {Rybarczyk, T. and Peaudecerf, B. and Penasa, M. and Gerlich, S. and Julsgaard, B. and M{\o}lmer, K. and Gleyzes, S. and Brune, M. and Raimond, J. M. and Haroche, S. and Dotsenko, I.},
    number = {6},
    pages = {1--9},
    volume = {91},
    doi = {10.1103/PhysRevA.91.062116},
    issn = {10941622}
}

@article{Bagrets2003,
    title = {{Full counting statistics of charge transfer in Coulomb blockade systems}},
    year = {2003},
    journal = {Physical Review B - Condensed Matter and Materials Physics},
    author = {Bagrets, D. A. and Nazarov, Yu V.},
    number = {8},
    pages = {1--16},
    volume = {67},
    doi = {10.1103/PhysRevB.67.085316},
    issn = {1550235X},
    arxivId = {cond-mat/0207624}
}

@article{Flindt2005,
    title = {{Full counting statistics of nano-electromechanical systems}},
    year = {2005},
    journal = {Europhysics Letters},
    author = {Flindt, C. and Novotn{\'{y}}, T. and Jauho, A. P.},
    number = {3},
    pages = {475--481},
    volume = {69},
    doi = {10.1209/epl/i2004-10351-x},
    issn = {02955075},
    arxivId = {cond-mat/0410322}
}

@article{Haase2018,
    title = {{Fundamental limits to frequency estimation: a comprehensive microscopic perspective}},
    year = {2018},
    journal = {New Journal of Physics},
    author = {Haase, J. F. and Smirne, A. and Ko{\l}ody{\'{n}}ski, J. and Demkowicz-Dobrza{\'{n}}ski, R. and Huelga, S. F.},
    number = {5},
    month = {5},
    pages = {053009},
    volume = {20},
    url = {https://iopscience.iop.org/article/10.1088/1367-2630/aab67f},
    doi = {10.1088/1367-2630/aab67f},
    issn = {1367-2630},
    arxivId = {1710.04673}
}

@article{Wimann2015,
    title = {{Generalized trace-distance measure connecting quantum and classical non-Markovianity}},
    year = {2015},
    journal = {Physical Review A},
    author = {Wi{\ss}mann, Steffen and Breuer, Heinz-Peter and Vacchini, Bassano},
    number = {4},
    month = {10},
    pages = {042108},
    volume = {92},
    url = {https://link.aps.org/doi/10.1103/PhysRevA.92.042108},
    doi = {10.1103/PhysRevA.92.042108},
    issn = {1050-2947},
    arxivId = {1507.08867}
}

@article{Settimo-RO,
    title = {{Generalized-rate-operator quantum jumps via realization-dependent transformations}},
    year = {2024},
    journal = {Physical Review A},
    author = {Settimo, Federico and Luoma, Kimmo and Chru{\'{s}}ci{\'{n}}ski, Dariusz and Vacchini, Bassano and Smirne, Andrea and Piilo, Jyrki},
    number = {6},
    month = {6},
    pages = {062201},
    volume = {109},
    url = {http://arxiv.org/abs/2402.12445 http://dx.doi.org/10.1103/PhysRevA.109.062201 https://link.aps.org/doi/10.1103/PhysRevA.109.062201},
    doi = {10.1103/PhysRevA.109.062201},
    issn = {2469-9926},
    arxivId = {2402.12445}
}

@article{Breuer2004,
    title = {{Genuine quantum trajectories for non-Markovian processes}},
    year = {2004},
    journal = {Physical Review A - Atomic, Molecular, and Optical Physics},
    author = {Breuer, Heinz Peter},
    number = {1},
    pages = {1--12},
    volume = {70},
    doi = {10.1103/PhysRevA.70.012106},
    issn = {10502947},
    arxivId = {quant-ph/0403117}
}

@book{Bengtsson2006,
    title = {{Geometry of Quantum States}},
    year = {2006},
    author = {Bengtsson, Ingemar and Zyczkowski, Karol},
    month = {5},
    publisher = {Cambridge University Press},
    url = {https://www.cambridge.org/core/product/identifier/9780511535048/type/book},
    isbn = {9780521814515},
    doi = {10.1017/CBO9780511535048}
}

@article{Breuer1998HeisenbergUnraveling,
    title = {{Heisenberg picture operators in the stochastic wave function approach to open quantum systems}},
    year = {1998},
    journal = {European Physical Journal D},
    author = {Breuer, H. P. and Kappler, B. and Petruccione, F.},
    number = {1},
    pages = {9--13},
    volume = {1},
    doi = {10.1007/s100530050058},
    issn = {14346060},
    arxivId = {quant-ph/9807080}
}

@article{Smirne2022,
    title = {{Holevo skew divergence for the characterization of information backflow}},
    year = {2022},
    journal = {Physical Review A},
    author = {Smirne, Andrea and Megier, Nina and Vacchini, Bassano},
    number = {1},
    month = {7},
    pages = {012205},
    volume = {106},
    url = {http://arxiv.org/abs/2201.07812 https://link.aps.org/doi/10.1103/PhysRevA.106.012205},
    doi = {10.1103/PhysRevA.106.012205},
    issn = {2469-9926},
    arxivId = {2201.07812}
}

@article{Minganti2020,
    title = {{Hybrid-Liouvillian formalism connecting exceptional points of non-Hermitian Hamiltonians and Liouvillians via postselection of quantum trajectories}},
    year = {2020},
    journal = {Physical Review A},
    author = {Minganti, Fabrizio and Miranowicz, Adam and Chhajlany, Ravindra W and Arkhipov, Ievgen I and Nori, Franco},
    number = {6},
    month = {6},
    pages = {062112},
    volume = {101},
    publisher = {American Physical Society},
    url = {https://link.aps.org/doi/10.1103/PhysRevA.101.062112},
    doi = {10.1103/PhysRevA.101.062112},
    issn = {2469-9926},
    arxivId = {2002.11620}
}

@article{Colla2022,
    title = {{Initial correlations in open quantum systems: constructing linear dynamical maps and master equations}},
    year = {2022},
    journal = {New Journal of Physics},
    author = {Colla, Alessandra and Neubrand, Niklas and Breuer, Heinz Peter},
    number = {12},
    volume = {24},
    doi = {10.1088/1367-2630/aca709},
    issn = {13672630},
    arxivId = {2210.13241}
}

@article{Gambetta2003,
    title = {{Interpretation of non-Markovian stochastic Schr{\"{o}}dinger equations as a hidden-variable theory}},
    year = {2003},
    journal = {Physical Review A - Atomic, Molecular, and Optical Physics},
    author = {Gambetta, Jay and Wiseman, H. M.},
    number = {6},
    pages = {9},
    volume = {68},
    doi = {10.1103/PhysRevA.68.062104},
    issn = {10941622},
    arxivId = {quant-ph/0307078}
}

@article{Gambetta-jump-nM,
    title = {{Jumplike unravelings for non-Markovian open quantum systems}},
    year = {2004},
    journal = {Physical Review A},
    author = {Gambetta, Jay and Askerud, T. and Wiseman, H. M.},
    number = {5},
    month = {5},
    pages = {052104},
    volume = {69},
    url = {https://link.aps.org/doi/10.1103/PhysRevA.69.052104},
    doi = {10.1103/PhysRevA.69.052104},
    issn = {1050-2947}
}

@article{Backer2024,
    title = {{Local Disclosure of Quantum Memory in Non-Markovian Dynamics}},
    year = {2024},
    journal = {Physical Review Letters},
    author = {B{\"{a}}cker, Charlotte and Beyer, Konstantin and Strunz, Walter T.},
    number = {6},
    month = {2},
    pages = {060402},
    volume = {132},
    url = {http://arxiv.org/abs/2310.01205 http://dx.doi.org/10.1103/PhysRevLett.132.060402 https://link.aps.org/doi/10.1103/PhysRevLett.132.060402},
    doi = {10.1103/PhysRevLett.132.060402},
    issn = {0031-9007},
    arxivId = {2310.01205}
}

@article{BLP-PRL,
    title = {{Measure for the Degree of Non-Markovian Behavior of Quantum Processes in Open Systems}},
    year = {2009},
    journal = {Physical Review Letters},
    author = {Breuer, Heinz Peter and Laine, Elsi Mari and Piilo, Jyrki},
    number = {21},
    pages = {1--4},
    volume = {103},
    doi = {10.1103/PhysRevLett.103.210401},
    issn = {00319007},
    arxivId = {0908.0238}
}

@article{BLP-PRA,
    title = {{Measure for the non-Markovianity of quantum processes}},
    year = {2010},
    journal = {Physical Review A},
    author = {Laine, Elsi-Mari and Piilo, Jyrki and Breuer, Heinz-Peter},
    number = {6},
    month = {6},
    pages = {062115},
    volume = {81},
    url = {https://link.aps.org/doi/10.1103/PhysRevA.81.062115},
    doi = {10.1103/PhysRevA.81.062115},
    issn = {1050-2947},
    arxivId = {1002.2583}
}

@article{Viola1997,
    title = {{Measured quantum dynamics of a trapped ion}},
    year = {1997},
    journal = {Physical Review A - Atomic, Molecular, and Optical Physics},
    author = {Viola, Lorenza and Onofrio, Roberto},
    number = {5},
    pages = {R3291-R3294},
    volume = {55},
    doi = {10.1103/PhysRevA.55.R3291},
    issn = {10941622}
}

@article{Barchielli1991,
    title = {{Measurements continuous in time and a posteriori states in quantum mechanics}},
    year = {1991},
    journal = {Journal of Physics A: General Physics},
    author = {Barchielli, A. and Belavkin, V. P.},
    number = {7},
    pages = {1495--1514},
    volume = {24},
    doi = {10.1088/0305-4470/24/7/022},
    issn = {03054470}
}

@article{Chruscinski2011,
    title = {{Measures of non-Markovianity: Divisibility versus backflow of information}},
    year = {2011},
    journal = {Physical Review A},
    author = {Chru{\'{s}}ci{\'{n}}ski, Dariusz and Kossakowski, Andrzej and Rivas, Ángel},
    number = {5},
    month = {5},
    pages = {052128},
    volume = {83},
    url = {https://link.aps.org/doi/10.1103/PhysRevA.83.052128},
    doi = {10.1103/PhysRevA.83.052128},
    issn = {1050-2947},
    arxivId = {1102.4318}
}

@article{Dum1992,
    title = {{Monte Carlo simulation of master equations in quantum optics for vacuum, thermal, and squeezed reservoirs}},
    year = {1992},
    journal = {Physical Review A},
    author = {Dum, R. and Parkins, A. S. and Zoller, P. and Gardiner, C. W.},
    number = {7},
    month = {10},
    pages = {4382--4396},
    volume = {46},
    url = {https://link.aps.org/doi/10.1103/PhysRevA.46.4382},
    doi = {10.1103/PhysRevA.46.4382},
    issn = {1050-2947}
}

@article{Piilo2008,
    title = {{Non-Markovian Quantum Jumps}},
    year = {2008},
    journal = {Physical Review Letters},
    author = {Piilo, Jyrki and Maniscalco, Sabrina and H{\"{a}}rk{\"{o}}nen, Kari and Suominen, Kalle-Antti},
    number = {18},
    month = {5},
    pages = {180402},
    volume = {100},
    url = {https://link.aps.org/doi/10.1103/PhysRevLett.100.180402},
    doi = {10.1103/PhysRevLett.100.180402},
    issn = {0031-9007},
    arxivId = {0706.4438}
}

@article{Ai2014AnFields,
    title = {{An efficient quantum jump method for coherent energy transfer dynamics in photosynthetic systems under the influence of laser fields}},
    year = {2014},
    journal = {New Journal of Physics},
    author = {Ai, Qing and Fan, Yuan Jia and Jin, Bih Yaw and Cheng, Yuan Chung},
    volume = {16},
    doi = {10.1088/1367-2630/16/5/053033},
    issn = {13672630},
    arxivId = {1404.2052}
}

@article{Li2018ConceptsHierarchy,
    title = {{Concepts of quantum non-Markovianity: A hierarchy}},
    year = {2018},
    journal = {Physics Reports},
    author = {Li, Li and Hall, Michael J.W. and Wiseman, Howard M.},
    pages = {1--51},
    volume = {759},
    doi = {10.1016/j.physrep.2018.07.001},
    issn = {03701573},
    arxivId = {1712.08879}
}

@article{Luoma2011ConnectingSystems,
    title = {{Connecting two jumplike unravelings for non-Markovian open quantum systems}},
    year = {2011},
    journal = {Physical Review A - Atomic, Molecular, and Optical Physics},
    author = {Luoma, Kimmo and Suominen, Kalle Antti and Piilo, Jyrki},
    number = {3},
    volume = {84},
    doi = {10.1103/PhysRevA.84.032113},
    issn = {10502947},
    arxivId = {1106.3173}
}

@article{Garraway1997DecayReservoir,
    title = {{Decay of an atom coupled strongly to a reservoir}},
    year = {1997},
    journal = {Physical Review A - Atomic, Molecular, and Optical Physics},
    author = {Garraway, B. M.},
    number = {6},
    pages = {4636--4639},
    volume = {55},
    doi = {10.1103/PhysRevA.55.4636},
    issn = {10941622}
}

@article{Luoma2014DetectingMonitoring,
    title = {{Detecting non-Markovianity from continuous monitoring}},
    year = {2014},
    journal = {Physical Review A - Atomic, Molecular, and Optical Physics},
    author = {Luoma, Kimmo and Haikka, Pinja and Piilo, Jyrki},
    number = {5},
    pages = {1--6},
    volume = {90},
    doi = {10.1103/PhysRevA.90.054101},
    issn = {10941622},
    arxivId = {1409.4516}
}

@article{Leggio2013EntropyTrajectories,
    title = {{Entropy production and information fluctuations along quantum trajectories}},
    year = {2013},
    journal = {Physical Review A - Atomic, Molecular, and Optical Physics},
    author = {Leggio, B. and Napoli, A. and Messina, A. and Breuer, H. P.},
    number = {4},
    pages = {1--10},
    volume = {88},
    doi = {10.1103/PhysRevA.88.042111},
    issn = {10502947},
    arxivId = {1305.6733}
}

@article{Gammelmark2014FisherMeasurements,
    title = {{Fisher information and the quantum Cram{\'{e}}r-Rao sensitivity limit of continuous measurements}},
    year = {2014},
    journal = {Physical Review Letters},
    author = {Gammelmark, Søren and M{\o}lmer, Klaus},
    number = {17},
    pages = {1--5},
    volume = {112},
    doi = {10.1103/PhysRevLett.112.170401},
    issn = {10797114},
    arxivId = {1310.5802}
}

@article{Brown2025GaugeTrajectories,
    title = {{Gauge freedoms in unravelled quantum dynamics: When do different continuous measurements yield identical quantum trajectories?}},
    year = {2025},
    journal = {Quantum},
    author = {Brown, Calum A. and Macieszczak, Katarzyna and Jack, Robert L.},
    month = {7},
    pages = {1787},
    volume = {9},
    url = {https://quantum-journal.org/papers/q-2025-07-09-1787/},
    doi = {10.22331/q-2025-07-09-1787},
    issn = {2521-327X}
}

@article{Radaelli2024GillespieTrajectories,
    title = {{Gillespie algorithm for quantum jump trajectories}},
    year = {2024},
    journal = {Physical Review A},
    author = {Radaelli, Marco and Landi, Gabriel T. and Binder, Felix C.},
    number = {6},
    month = {12},
    pages = {062212},
    volume = {110},
    url = {http://arxiv.org/abs/2303.15405 http://dx.doi.org/10.1103/PhysRevA.110.062212 https://link.aps.org/doi/10.1103/PhysRevA.110.062212},
    doi = {10.1103/PhysRevA.110.062212},
    issn = {2469-9926},
    arxivId = {2303.15405}
}

@article{Steinbach1995High-orderEvolution,
    title = {{High-order unraveling of master equations for dissipative evolution}},
    year = {1995},
    journal = {Physical Review A},
    author = {Steinbach, J. and Garraway, B. M. and Knight, P. L.},
    number = {4},
    pages = {3302--3308},
    volume = {51},
    doi = {10.1103/PhysRevA.51.3302},
    issn = {10502947}
}

@article{Chruscinski2022HowRepresentations,
    title = {{How to design quantum-jump trajectories via distinct master equation representations}},
    year = {2022},
    journal = {Quantum},
    author = {Chru{\'{s}}ci{\'{n}}ski, Dariusz and Luoma, Kimmo and Piilo, Jyrki and Smirne, Andrea},
    pages = {1--20},
    volume = {6},
    doi = {10.22331/Q-2022-10-13-835},
    issn = {2521327X},
    arxivId = {2009.11312}
}

@book{Klebaner2012IntroductionApplications,
    title = {{Introduction to Stochastic Calculus with Applications}},
    year = {2012},
    author = {Klebaner, Fima C},
    month = {3},
    publisher = {IMPERIAL COLLEGE PRESS},
    url = {https://www.worldscientific.com/worldscibooks/10.1142/p821},
    isbn = {978-1-84816-831-2},
    doi = {10.1142/p821}
}

@article{Mlmer1993MonteOptics,
    title = {{Monte Carlo wave-function method in quantum optics}},
    year = {1993},
    journal = {Journal of the Optical Society of America B},
    author = {M{\o}lmer, Klaus and Castin, Yvan and Dalibard, Jean},
    number = {3},
    pages = {524},
    volume = {10},
    doi = {10.1364/josab.10.000524},
    issn = {0740-3224}
}

@article{Menczel2024Non-HermitianThermodynamics,
    title = {{Non-Hermitian pseudomodes for strongly coupled open quantum systems: Unravelings, correlations, and thermodynamics}},
    year = {2024},
    journal = {Physical Review Research},
    author = {Menczel, Paul and Funo, Ken and Cirio, Mauro and Lambert, Neill and Nori, Franco},
    number = {3},
    volume = {6},
    publisher = {American Physical Society},
    doi = {10.1103/PhysRevResearch.6.033237},
    issn = {26431564},
    arxivId = {2401.11830}
}

@article{Diosi2008Non-markovianObservables,
    title = {{Non-markovian continuous quantum measurement of retarded observables}},
    year = {2008},
    journal = {Physical Review Letters},
    author = {Di{\'{o}}si, Lajos},
    number = {8},
    pages = {1--4},
    volume = {100},
    doi = {10.1103/PhysRevLett.100.080401},
    issn = {10797114},
    arxivId = {0710.5489}
}

@article{Rebentrost2009,
    title = {{Non-Markovian quantum jumps in excitonic energy transfer}},
    year = {2009},
    journal = {Journal of Chemical Physics},
    author = {Rebentrost, Patrick and Chakraborty, Rupak and Aspuru-Guzik, Alán},
    number = {18},
    pages = {1--9},
    volume = {131},
    doi = {10.1063/1.3259838},
    issn = {00219606},
    arxivId = {0908.1961}
}

@article{Diosi-NMQSD,
    title = {{Non-Markovian quantum state diffusion}},
    year = {1998},
    journal = {Physical Review A - Atomic, Molecular, and Optical Physics},
    author = {Di{\'{o}}si, L. and Gisin, N. and Strunz, W. T.},
    number = {3},
    pages = {1699--1712},
    volume = {58},
    doi = {10.1103/PhysRevA.58.1699},
    issn = {10941622},
    arxivId = {quant-ph/9803062}
}

@article{Luoma2012,
    title = {{Non-Markovian waiting-time distribution for quantum jumps in open systems}},
    year = {2012},
    journal = {Physical Review A},
    author = {Luoma, Kimmo and H{\"{a}}rk{\"{o}}nen, Kari and Maniscalco, Sabrina and Suominen, Kalle-Antti and Piilo, Jyrki},
    number = {2},
    month = {8},
    pages = {022102},
    volume = {86},
    url = {https://link.aps.org/doi/10.1103/PhysRevA.86.022102},
    doi = {10.1103/PhysRevA.86.022102},
    issn = {1050-2947}
}

@article{Roch2014,
    title = {{Observation of measurement-induced entanglement and quantum trajectories of remote superconducting qubits}},
    year = {2014},
    journal = {Physical Review Letters},
    author = {Roch, N. and Schwartz, M. E. and Motzoi, F. and Macklin, C. and Vijay, R. and Eddins, A. W. and Korotkov, A. N. and Whaley, K. B. and Sarovar, M. and Siddiqi, I.},
    number = {17},
    pages = {1--5},
    volume = {112},
    doi = {10.1103/PhysRevLett.112.170501},
    issn = {10797114},
    arxivId = {1402.1868}
}

@article{Bergquist1986,
    title = {{Observation of quantum jumps in a single atom}},
    year = {1986},
    journal = {Physical Review Letters},
    author = {Bergquist, J. C. and Hulet, Randall G. and Itano, Wayne M. and Wineland, D. J.},
    number = {14},
    pages = {1699--1702},
    volume = {57},
    doi = {10.1103/PhysRevLett.57.1699},
    issn = {00319007},
    pmid = {10033522}
}

@article{Vijay2011,
    title = {{Observation of Quantum Jumps in a Superconducting Artificial Atom}},
    year = {2011},
    journal = {Physical Review Letters},
    author = {Vijay, R. and Slichter, D. H. and Siddiqi, I.},
    number = {11},
    month = {3},
    pages = {110502},
    volume = {106},
    url = {https://link.aps.org/doi/10.1103/PhysRevLett.106.110502},
    doi = {10.1103/PhysRevLett.106.110502},
    issn = {0031-9007}
}

@article{Campagne-Ibarcq2016,
    title = {{Observing quantum state diffusion by heterodyne detection of fluorescence}},
    year = {2016},
    journal = {Physical Review X},
    author = {Campagne-Ibarcq, P. and Six, P. and Bretheau, L. and Sarlette, A. and Mirrahimi, M. and Rouchon, P. and Huard, B.},
    number = {1},
    pages = {1--7},
    volume = {6},
    doi = {10.1103/PhysRevX.6.011002},
    issn = {21603308},
    arxivId = {1511.01415}
}

@article{Peil1999,
    title = {{Observing the Quantum Limit of an Electron Cyclotron: QND Measurements of Quantum Jumps between Fock States}},
    year = {1999},
    journal = {Physical Review Letters},
    author = {Peil, S. and Gabrielse, G.},
    number = {7},
    month = {8},
    pages = {1287--1290},
    volume = {83},
    url = {https://link.aps.org/doi/10.1103/PhysRevLett.83.1287},
    doi = {10.1103/PhysRevLett.83.1287},
    issn = {0031-9007}
}

@article{Lindblad1976,
    title = {{On the generators of quantum dynamical semigroups}},
    year = {1976},
    journal = {Communications in Mathematical Physics},
    author = {Lindblad, Goran},
    number = {2},
    month = {6},
    pages = {119--130},
    volume = {48},
    url = {http://link.springer.com/10.1007/BF01608499},
    doi = {10.1007/BF01608499},
    issn = {0010-3616}
}

@article{Smirne2022b,
    title = {{On the Use of Total State Decompositions for the Study of Reduced Dynamics}},
    year = {2022},
    journal = {Open Systems and Information Dynamics},
    author = {Smirne, Andrea and Megier, Nina and Vacchini, Bassano},
    number = {2},
    pages = {1--20},
    volume = {29},
    doi = {10.1142/S1230161222500081},
    issn = {12301612},
    arxivId = {2209.02288}
}

@book{Rivas-Huelga-OQS,
    title = {{Open Quantum Systems}},
    year = {2012},
    author = {Rivas, Ángel and Huelga, Susana F.},
    series = {SpringerBriefs in Physics},
    publisher = {Springer Berlin Heidelberg},
    url = {https://link.springer.com/10.1007/978-3-642-23354-8},
    address = {Berlin, Heidelberg},
    isbn = {978-3-642-23353-1},
    doi = {10.1007/978-3-642-23354-8}
}

@book{Vacchini-OQS,
    title = {{Open Quantum Systems}},
    year = {2024},
    author = {Vacchini, Bassano},
    series = {Graduate Texts in Physics},
    publisher = {Springer Nature Switzerland},
    url = {https://link.springer.com/10.1007/978-3-031-58218-9},
    address = {Cham},
    isbn = {978-3-031-58217-2},
    doi = {10.1007/978-3-031-58218-9}
}

@article{Modi2012,
    title = {{Operational approach to open dynamics and quantifying initial correlations}},
    year = {2012},
    journal = {Scientific Reports},
    author = {Modi, Kavan},
    volume = {2},
    doi = {10.1038/srep00581},
    issn = {20452322},
    arxivId = {1011.6138}
}

@article{Diosi-orthogonal-jumps,
    title = {{Orthogonal jumps of the wavefunction in white-noise potentials}},
    year = {1985},
    journal = {Physics Letters A},
    author = {Di{\'{o}}si, Lajos},
    number = {6-7},
    pages = {288--292},
    volume = {112},
    doi = {10.1016/0375-9601(85)90342-1},
    issn = {03759601},
    arxivId = {1501.00274}
}

@article{Radaelli2026,
    title = {{Parameter estimation for quantum jump unraveling}},
    year = {2026},
    journal = {Quantum},
    author = {Radaelli, Marco and Smiga, Joseph A. and Landi, Gabriel T. and Binder, Felix C.},
    month = {2},
    pages = {1993},
    volume = {10},
    url = {http://arxiv.org/abs/2402.06556 https://quantum-journal.org/papers/q-2026-02-02-1993/},
    doi = {10.22331/q-2026-02-02-1993},
    issn = {2521-327X},
    arxivId = {2402.06556}
}

@article{Filippov2020,
    title = {{Phase Covariant Qubit Dynamics and Divisibility}},
    year = {2020},
    journal = {Lobachevskii Journal of Mathematics},
    author = {Filippov, S. N. and Glinov, A. N. and Lepp{\"{a}}j{\"{a}}rvi, L.},
    number = {4},
    month = {4},
    pages = {617--630},
    volume = {41},
    url = {https://link.springer.com/10.1134/S1995080220040095},
    doi = {10.1134/S1995080220040095},
    issn = {1995-0802},
    arxivId = {1911.09468}
}

@article{Liu2013,
    title = {{Photonic realization of nonlocal memory effects and non-Markovian quantum probes}},
    year = {2013},
    journal = {Scientific Reports},
    author = {Liu, Bi Heng and Cao, Dong Yang and Huang, Yun Feng and Li, Chuan Feng and Guo, Guang Can and Laine, Elsi Mari and Breuer, Heinz Peter and Piilo, Jyrki},
    pages = {1--6},
    volume = {3},
    doi = {10.1038/srep01781},
    issn = {20452322},
    arxivId = {1208.1358}
}

@book{Alicki2007,
    title = {{Quantum Dynamical Semigroups and Applications}},
    year = {2007},
    author = {Alicki, Robert and Lendi, Karl},
    series = {Lecture Notes in Physics},
    volume = {717},
    publisher = {Springer Berlin Heidelberg},
    url = {http://link.springer.com/10.1007/3-540-70861-8},
    address = {Berlin, Heidelberg},
    isbn = {978-3-540-70860-5},
    doi = {10.1007/3-540-70861-8}
}

@article{Leibfried2003,
    title = {{Quantum dynamics of single trapped ions}},
    year = {2003},
    journal = {Reviews of Modern Physics},
    author = {Leibfried, D. and Blatt, R. and Monroe, C. and Wineland, D.},
    number = {1},
    pages = {281--324},
    volume = {75},
    doi = {10.1103/RevModPhys.75.281},
    issn = {00346861}
}

@article{Minganti2019,
    title = {{Quantum exceptional points of non-Hermitian Hamiltonians and Liouvillians: The effects of quantum jumps}},
    year = {2019},
    journal = {Physical Review A},
    author = {Minganti, Fabrizio and Miranowicz, Adam and Chhajlany, Ravindra W. and Nori, Franco},
    number = {6},
    volume = {100},
    doi = {10.1103/PhysRevA.100.062131},
    issn = {24699934},
    arxivId = {1909.11619}
}

@article{Zhang2017,
    title = {{Quantum feedback: Theory, experiments, and applications}},
    year = {2017},
    journal = {Physics Reports},
    author = {Zhang, Jing and Liu, Yu xi and Wu, Re Bing and Jacobs, Kurt and Nori, Franco},
    pages = {1--60},
    volume = {679},
    publisher = {Elsevier B.V.},
    url = {http://dx.doi.org/10.1016/j.physrep.2017.02.003},
    doi = {10.1016/j.physrep.2017.02.003},
    issn = {03701573},
    arxivId = {1407.8536}
}

@article{Gleyzes2007,
    title = {{Quantum jumps of light recording the birth and death of a photon in a cavity}},
    year = {2007},
    journal = {Nature},
    author = {Gleyzes, Sébastien and Kuhr, Stefan and Guerlin, Christine and Bernu, Julien and Del{\'{e}}glise, Samuel and Busk Hoff, Ulrich and Brune, Michel and Raimond, Jean Michel and Haroche, Serge},
    number = {7133},
    pages = {297--300},
    volume = {446},
    doi = {10.1038/nature05589},
    issn = {14764687}
}

@article{Budini2022,
    title = {{Quantum Non-Markovian Environment-to-System Backflows of Information: Nonoperational vs. Operational Approaches}},
    year = {2022},
    journal = {Entropy},
    author = {Budini, Adrián A},
    number = {5},
    pages = {1--17},
    volume = {24},
    doi = {10.3390/e24050649},
    issn = {10994300},
    arxivId = {2205.03333}
}

@article{rivas-quantum-nm,
    title = {{Quantum non-Markovianity: characterization, quantification and detection}},
    year = {2014},
    journal = {Reports on Progress in Physics},
    author = {Rivas, Ángel and Huelga, Susana F. and Plenio, Martin B.},
    number = {9},
    month = {9},
    pages = {094001},
    volume = {77},
    publisher = {IOP Publishing},
    url = {https://iopscience-iop-org.pros1.lib.unimi.it/article/10.1088/0034-4885/77/9/094001 https://iopscience-iop-org.pros1.lib.unimi.it/article/10.1088/0034-4885/77/9/094001/meta https://iopscience.iop.org/article/10.1088/0034-4885/77/9/094001},
    doi = {10.1088/0034-4885/77/9/094001},
    issn = {0034-4885},
    arxivId = {1405.0303}
}

@book{Percival1999,
    title = {{Quantum State Diffusion}},
    year = {1998},
    author = {Percival, Ian},
    pages = {198},
    publisher = {Cambridge University Press},
    isbn = {9780521620079}
}

@article{Wiseman1994,
    title = {{Quantum theory of continuous feedback}},
    year = {1994},
    journal = {Physical Review A},
    author = {Wiseman, H. M.},
    number = {3},
    pages = {2133--2150},
    volume = {49},
    doi = {10.1103/PhysRevA.49.2133},
    issn = {10502947}
}

@article{Wiseman1993,
    title = {{Quantum theory of optical feedback via homodyne detection}},
    year = {1993},
    journal = {Physical Review Letters},
    author = {Wiseman, H. M. and Milburn, G. J.},
    number = {5},
    pages = {548--551},
    volume = {70},
    doi = {10.1103/PhysRevLett.70.548},
    issn = {00319007}
}

@book{Barchielli2009,
    title = {{Quantum Trajectories and Measurements in Continuous Time}},
    year = {2009},
    author = {Barchielli, Alberto and Gregoratti, Matteo},
    series = {Lecture Notes in Physics},
    volume = {782},
    publisher = {Springer Berlin Heidelberg},
    url = {https://link.springer.com/10.1007/978-3-642-01298-3},
    address = {Berlin, Heidelberg},
    isbn = {978-3-642-01297-6},
    doi = {10.1007/978-3-642-01298-3}
}

@article{Becker2023,
    title = {{Quantum Trajectories for Time-Local Non-Lindblad Master Equations}},
    year = {2023},
    journal = {Physical Review Letters},
    author = {Becker, Tobias and Netzer, Ché and Eckardt, André},
    number = {16},
    month = {10},
    pages = {160401},
    volume = {131},
    url = {http://arxiv.org/abs/2306.14876 http://dx.doi.org/10.1103/PhysRevLett.131.160401 https://link.aps.org/doi/10.1103/PhysRevLett.131.160401},
    doi = {10.1103/PhysRevLett.131.160401},
    issn = {0031-9007},
    arxivId = {2306.14876}
}

@article{Weber2016,
    title = {{Quantum trajectories of superconducting qubits}},
    year = {2016},
    journal = {Comptes Rendus Physique},
    author = {Weber, Steven J. and Murch, Kater W. and Kimchi-Schwartz, Mollie E. and Roch, Nicolas and Siddiqi, Irfan},
    number = {7},
    pages = {766--777},
    volume = {17},
    publisher = {Elsevier Masson SAS},
    url = {http://dx.doi.org/10.1016/j.crhy.2016.07.007},
    doi = {10.1016/j.crhy.2016.07.007},
    issn = {18781535},
    arxivId = {1506.08165}
}

@article{Gambetta2008,
    title = {{Quantum trajectory approach to circuit QED: Quantum jumps and the Zeno effect}},
    year = {2008},
    journal = {Physical Review A - Atomic, Molecular, and Optical Physics},
    author = {Gambetta, Jay and Blais, Alexandre and Boissonneault, M. and Houck, A. A. and Schuster, D. I. and Girvin, S. M.},
    number = {1},
    pages = {1--18},
    volume = {77},
    doi = {10.1103/PhysRevA.77.012112},
    issn = {10502947},
    arxivId = {0709.4264}
}

@article{Donvil2022,
    title = {{Quantum trajectory framework for general time-local master equations.}},
    year = {2022},
    journal = {Nature communications},
    author = {Donvil, Brecht and Muratore-Ginanneschi, Paolo},
    number = {1},
    month = {7},
    pages = {4140},
    volume = {13},
    publisher = {Springer US},
    url = {https://www.nature.com/articles/s41467-022-31533-8 http://www.ncbi.nlm.nih.gov/pubmed/35842427 http://www.pubmedcentral.nih.gov/articlerender.fcgi?artid=PMC9288492},
    doi = {10.1038/s41467-022-31533-8},
    issn = {2041-1723},
    pmid = {35842427}
}

@article{Garrahan2011,
    title = {{Quantum trajectory phase transitions in the micromaser}},
    year = {2011},
    journal = {Physical Review E - Statistical, Nonlinear, and Soft Matter Physics},
    author = {Garrahan, Juan P. and Armour, Andrew D. and Lesanovsky, Igor},
    number = {2},
    pages = {1--6},
    volume = {84},
    doi = {10.1103/PhysRevE.84.021115},
    issn = {15393755},
    arxivId = {1103.0919}
}

@article{Ozawa2018,
    title = {{Quantum Zeno Effect assisted Spectroscopy of a single trapped Ion}},
    year = {2018},
    journal = {Scientific Reports},
    author = {Ozawa, Akira and Davila-Rodriguez, Josue and H{\"{a}}nsch, Theodor W. and Udem, Thomas},
    number = {1},
    pages = {1--8},
    volume = {8},
    doi = {10.1038/s41598-018-28824-w}
}

@article{qutip2,
    title = {{QuTiP 2: A Python framework for the dynamics of open quantum systems}},
    year = {2013},
    journal = {Computer Physics Communications},
    author = {Johansson, J.R. and Nation, P.D. and Nori, Franco},
    number = {4},
    month = {4},
    pages = {1234--1240},
    volume = {184},
    url = {https://linkinghub.elsevier.com/retrieve/pii/S0010465512003955},
    doi = {10.1016/j.cpc.2012.11.019},
    issn = {00104655}
}

@article{qutip,
    title = {{QuTiP: An open-source Python framework for the dynamics of open quantum systems}},
    year = {2012},
    journal = {Computer Physics Communications},
    author = {Johansson, J.R. and Nation, P.D. and Nori, Franco},
    number = {8},
    month = {8},
    pages = {1760--1772},
    volume = {183},
    url = {https://linkinghub.elsevier.com/retrieve/pii/S0010465512000835},
    doi = {10.1016/j.cpc.2012.02.021},
    issn = {00104655}
}

@article{Smirne2020,
    title = {{Rate Operator Unraveling for Open Quantum System Dynamics}},
    year = {2020},
    journal = {Physical Review Letters},
    author = {Smirne, Andrea and Caiaffa, Matteo and Piilo, Jyrki},
    number = {19},
    pages = {190402},
    volume = {124},
    publisher = {American Physical Society},
    url = {https://doi.org/10.1103/PhysRevLett.124.190402},
    doi = {10.1103/PhysRevLett.124.190402},
    issn = {10797114},
    pmid = {32469534},
    arxivId = {2004.09537}
}

@article{Sayrin2011,
    title = {{Real-time quantum feedback prepares and stabilizes photon number states}},
    year = {2011},
    journal = {Nature},
    author = {Sayrin, Clément and Dotsenko, Igor and Zhou, Xingxing and Peaudecerf, Bruno and Rybarczyk, Théo and Gleyzes, Sébastien and Rouchon, Pierre and Mirrahimi, Mazyar and Amini, Hadis and Brune, Michel and Raimond, Jean Michel and Haroche, Serge},
    number = {7362},
    pages = {73--77},
    volume = {477},
    doi = {10.1038/nature10376},
    issn = {00280836}
}

@article{Vacchini2016,
    title = {{Reduced dynamical maps in the presence of initial correlations}},
    year = {2016},
    journal = {Scientific Reports},
    author = {Vacchini, Bassano and Amato, Giulio},
    number = {November},
    pages = {1--12},
    volume = {6},
    publisher = {Nature Publishing Group},
    url = {http://dx.doi.org/10.1038/srep37328},
    doi = {10.1038/srep37328},
    issn = {20452322}
}

@article{Pechukas1994,
    title = {{Reduced dynamics need not be completely positive}},
    year = {1994},
    journal = {Physical Review Letters},
    author = {Pechukas, Philip},
    number = {8},
    pages = {1060--1062},
    volume = {73},
    doi = {10.1103/PhysRevLett.73.1060},
    issn = {00319007}
}

@article{Teittinen2018,
    title = {{Revealing memory effects in phase-covariant quantum master equations}},
    year = {2018},
    journal = {New Journal of Physics},
    author = {Teittinen, J and Lyyra, H and Sokolov, B and Maniscalco, S},
    number = {7},
    month = {7},
    pages = {073012},
    volume = {20},
    url = {https://iopscience.iop.org/article/10.1088/1367-2630/aacc38},
    doi = {10.1088/1367-2630/aacc38},
    issn = {1367-2630}
}

@article{Jelezko2002,
    title = {{Single spin states in a defect center resolved by optical spectroscopy}},
    year = {2002},
    journal = {Applied Physics Letters},
    author = {Jelezko, F. and Popa, I. and Gruber, A. and Tietz, C. and Wrachtrup, J. and Nizovtsev, A. and Kilin, S.},
    number = {12},
    month = {9},
    pages = {2160--2162},
    volume = {81},
    url = {https://pubs.aip.org/apl/article/81/12/2160/114736/Single-spin-states-in-a-defect-center-resolved-by},
    doi = {10.1063/1.1507838},
    issn = {0003-6951}
}

@article{Bao2020,
    title = {{Spin squeezing of 1011 atoms by prediction and retrodiction measurements}},
    year = {2020},
    journal = {Nature},
    author = {Bao, Han and Duan, Junlei and Jin, Shenchao and Lu, Xingda and Li, Pengxiong and Qu, Weizhi and Wang, Mingfeng and Novikova, Irina and Mikhailov, Eugeniy E. and Zhao, Kai Feng and M{\o}lmer, Klaus and Shen, Heng and Xiao, Yanhong},
    number = {7807},
    pages = {159--163},
    volume = {581},
    publisher = {Springer US},
    url = {http://dx.doi.org/10.1038/s41586-020-2243-7},
    doi = {10.1038/s41586-020-2243-7},
    issn = {14764687},
    pmid = {32405021}
}

@article{Dean2009,
    title = {{Splitting for rare event simulation: A large deviation approach to design and analysis}},
    year = {2009},
    journal = {Stochastic Processes and their Applications},
    author = {Dean, Thomas and Dupuis, Paul},
    number = {2},
    pages = {562--587},
    volume = {119},
    publisher = {Elsevier B.V.},
    doi = {10.1016/j.spa.2008.02.017},
    issn = {03044149}
}

@article{Breuer2008,
    title = {{Stochastic jump processes for non-Markovian quantum dynamics}},
    year = {2009},
    journal = {EPL (Europhysics Letters)},
    author = {Breuer, H.-P. and Piilo, Jyrki},
    number = {5},
    month = {3},
    pages = {50004},
    volume = {85},
    url = {http://arxiv.org/abs/0810.5511 http://dx.doi.org/10.1209/0295-5075/85/50004 https://iopscience.iop.org/article/10.1209/0295-5075/85/50004},
    doi = {10.1209/0295-5075/85/50004},
    issn = {0295-5075},
    arxivId = {0810.5511}
}

@article{Diosi-stochastic-repr,
    title = {{Stochastic pure state representation for open quantum systems}},
    year = {1986},
    journal = {Physics Letters A},
    author = {Di{\'{o}}si, Lajos},
    number = {8-9},
    pages = {451--454},
    volume = {114},
    doi = {10.1016/0375-9601(86)90692-4},
    issn = {03759601},
    arxivId = {1609.09636}
}

@article{Caiaffa-W-diffusive,
    title = {{Stochastic unraveling of positive quantum dynamics}},
    year = {2017},
    journal = {Physical Review A},
    author = {Caiaffa, Matteo and Smirne, Andrea and Bassi, Angelo},
    number = {6},
    month = {6},
    pages = {062101},
    volume = {95},
    url = {http://link.aps.org/doi/10.1103/PhysRevA.95.062101},
    doi = {10.1103/PhysRevA.95.062101},
    issn = {2469-9926},
    arxivId = {1612.04546}
}

@article{Settimo2026,
    title = {{Stochastic unravelings for Heisenberg picture and trace-nonpreserving dynamics}},
    year = {2026},
    journal = {Physical Review A},
    author = {Settimo, Federico and Luoma, Kimmo and Chru{\'{s}}ci{\'{n}}ski, Dariusz and Vacchini, Bassano and Smirne, Andrea and Piilo, Jyrki},
    number = {4},
    month = {4},
    pages = {042444},
    volume = {113},
    url = {https://link.aps.org/doi/10.1103/hwfw-2l3c},
    doi = {10.1103/hwfw-2l3c},
    issn = {2469-9926}
}

@article{Breuer1998Doubled,
    title = {{Stochastic wave function approach to generalized master equations}},
    year = {1998},
    journal = {Journal of Superconductivity},
    author = {Breuer, H. P. and Kappler, B. and Petruccione, F.},
    month = {6},
    pages = {695–702},
    volume = {12},
    url = {http://arxiv.org/abs/quant-ph/9806026},
    arxivId = {quant-ph/9806026}
}

@article{Imamoglu1994,
    title = {{Stochastic wave-function approach to non-Markovian systems}},
    year = {1994},
    journal = {Physical Review A},
    author = {Imamoglu, A.},
    number = {5},
    pages = {3650--3653},
    volume = {50},
    doi = {10.1103/PhysRevA.50.3650},
    issn = {10502947}
}

@article{Breuer1999,
    title = {{Stochastic wave-function method for non-Markovian quantum master equations}},
    year = {1999},
    journal = {Physical Review A - Atomic, Molecular, and Optical Physics},
    author = {Breuer, Heinz Peter and Kappler, Bernd and Petruccione, Francesco},
    number = {2},
    pages = {1633--1643},
    volume = {59},
    doi = {10.1103/PhysRevA.59.1633},
    issn = {10941622},
    arxivId = {quant-ph/9906024}
}

@article{Plenio1998,
    title = {{The quantum-jump approach to dissipative dynamics in quantum optics}},
    year = {1998},
    journal = {Reviews of Modern Physics},
    author = {Plenio, M. B. and Knight, P. L.},
    number = {1},
    month = {1},
    pages = {101--144},
    volume = {70},
    url = {https://link.aps.org/doi/10.1103/RevModPhys.70.101},
    doi = {10.1103/RevModPhys.70.101},
    issn = {0034-6861}
}

@article{Gisin1992,
    title = {{The quantum-state diffusion model applied to open systems}},
    year = {1992},
    journal = {Journal of Physics A: General Physics},
    author = {Gisin, N. and Percival, I. C.},
    number = {21},
    pages = {5677--5691},
    volume = {25},
    doi = {10.1088/0305-4470/25/21/023},
    issn = {03054470}
}

@book{Breuer-Petruccione,
    title = {{The Theory of Open Quantum Systems}},
    year = {2007},
    author = {Breuer, Heinz-Peter and Petruccione, Francesco},
    month = {1},
    publisher = {Oxford University PressOxford},
    url = {https://academic.oup.com/book/27757},
    isbn = {0199213909},
    doi = {10.1093/acprof:oso/9780199213900.001.0001}
}

@article{Garrahan2010,
    title = {{Thermodynamics of quantum jump trajectories}},
    year = {2010},
    journal = {Physical Review Letters},
    author = {Garrahan, Juan P and Lesanovsky, Igor},
    number = {16},
    pages = {1--4},
    volume = {104},
    doi = {10.1103/PhysRevLett.104.160601},
    issn = {00319007},
    arxivId = {0911.0556}
}

@article{Perfetto2022,
    title = {{Thermodynamics of quantum-jump trajectories of open quantum systems subject to stochastic resetting}},
    year = {2022},
    journal = {SciPost Physics},
    author = {Perfetto, Gabriele and Carollo, Federico and Lesanovsky, Igor},
    number = {4},
    month = {10},
    pages = {079},
    volume = {13},
    url = {https://scipost.org/10.21468/SciPostPhys.13.4.079},
    doi = {10.21468/SciPostPhys.13.4.079},
    issn = {2542-4653}
}

@article{Minev2019,
    title = {{To catch and reverse a quantum jump mid-flight}},
    year = {2019},
    journal = {Nature},
    author = {Minev, Z. K. and Mundhada, S. O. and Shankar, S. and Reinhold, P. and Guti{\'{e}}rrez-J{\'{a}}uregui, R. and Schoelkopf, R. J. and Mirrahimi, M. and Carmichael, H. J. and Devoret, M. H.},
    number = {7760},
    pages = {200--204},
    volume = {570},
    doi = {10.1038/s41586-019-1287-z},
    issn = {14764687},
    pmid = {31160725},
    arxivId = {1803.00545}
}

@article{Cavina2025,
    title = {{Unifying quantum stochastic methods using Wick's theorem on the Keldysh contour}},
    year = {2025},
    journal = {Physical Review Research},
    author = {Cavina, Vasco and D’Abbruzzo, Antonio and Giovannetti, Vittorio},
    number = {4},
    month = {12},
    pages = {043262},
    volume = {7},
    url = {https://link.aps.org/doi/10.1103/cnvm-w8cy},
    doi = {10.1103/cnvm-w8cy},
    issn = {2643-1564}
}

@article{Gardiner1992,
    title = {{Wave-function quantum stochastic differential equations and quantum-jump simulation methods}},
    year = {1992},
    journal = {Physical Review A},
    author = {Gardiner, C. W. and Parkins, A. S. and Zoller, P.},
    number = {7},
    pages = {4363--4381},
    volume = {46},
    doi = {10.1103/PhysRevA.46.4363},
    issn = {10502947}
}

@article{Shaji2005,
    title = {{Who's afraid of not completely positive maps?}},
    year = {2005},
    journal = {Physics Letters, Section A: General, Atomic and Solid State Physics},
    author = {Shaji, Anil and Sudarshan, E. C.G.},
    number = {1-4},
    pages = {48--54},
    volume = {341},
    doi = {10.1016/j.physleta.2005.04.029},
    issn = {03759601}
}

@article{Gambetta2002Non-MarkovianTheory,
    title = {{Non-Markovian stochastic Schr{\"{o}}dinger equations: Generalization to real-valued noise using quantum-measurement theory}},
    year = {2002},
    journal = {Physical Review A - Atomic, Molecular, and Optical Physics},
    author = {Gambetta, Jay and Wiseman, H. M.},
    number = {1},
    pages = {17},
    volume = {66},
    doi = {10.1103/PhysRevA.66.012108},
    issn = {10941622},
    arxivId = {quant-ph/0202117}
}

@article{Garraway1997NonperturbativeCavity,
    title = {{Nonperturbative decay of an atomic system in a cavity}},
    year = {1997},
    journal = {Physical Review A - Atomic, Molecular, and Optical Physics},
    author = {Garraway, B M},
    number = {3},
    pages = {2290--2303},
    volume = {55},
    doi = {10.1103/PhysRevA.55.2290},
    issn = {10941622}
}

@article{Kossakowski1972OnSemigroup,
    title = {{On necessary and sufficient conditions for a generator of a quantum dynamical semigroup}},
    year = {1972},
    journal = {Bull. Acad. Sci. Math},
    author = {Kossakowski, Andrzej},
    pages = {1021},
    volume = {20}
}

@article{Piilo2009OpenJumps,
    title = {{Open system dynamics with non-Markovian quantum jumps}},
    year = {2009},
    journal = {Physical Review A - Atomic, Molecular, and Optical Physics},
    author = {Piilo, J. and H{\"{a}}rk{\"{o}}nen, K. and Maniscalco, S. and Suominen, K. A.},
    number = {6},
    pages = {1--17},
    volume = {79},
    doi = {10.1103/PhysRevA.79.062112},
    issn = {10502947},
    arxivId = {0902.3609}
}

@article{Wimann2012OptimalDynamics,
    title = {{Optimal state pairs for non-Markovian quantum dynamics}},
    year = {2012},
    journal = {Physical Review A - Atomic, Molecular, and Optical Physics},
    author = {Wimann, Steffen and Karlsson, Antti and Laine, Elsi Mari and Piilo, Jyrki and Breuer, Heinz Peter},
    number = {6},
    month = {12},
    volume = {86},
    doi = {10.1103/PhysRevA.86.062108},
    issn = {10502947},
    arxivId = {1209.4989}
}

@book{Williams1991ProbabilityMartingales,
    title = {{Probability with Martingales}},
    year = {1991},
    author = {Williams, David},
    month = {2},
    publisher = {Cambridge University Press},
    url = {https://www.cambridge.org/core/product/identifier/9780511813658/type/book},
    isbn = {9780521404556},
    doi = {10.1017/CBO9780511813658}
}

@article{Wiseman2008Pure-stateExist,
    title = {{Pure-state quantum trajectories for general non-markovian systems do not exist}},
    year = {2008},
    journal = {Physical Review Letters},
    author = {Wiseman, Howard M. and Gambetta, J. M.},
    number = {14},
    pages = {1--4},
    volume = {101},
    doi = {10.1103/PhysRevLett.101.140401},
    issn = {00319007},
    pmid = {18851507}
}

@book{Wiseman2009QuantumControl,
    title = {{Quantum Measurement and Control}},
    year = {2009},
    author = {Wiseman, Howard M. and Milburn, Gerard J.},
    month = {11},
    publisher = {Cambridge University Press},
    url = {https://www.cambridge.org/core/product/identifier/9780511813948/type/book},
    isbn = {9780521804424},
    doi = {10.1017/CBO9780511813948}
}

@book{Gardiner2004QuantumOptics,
    title = {{Quantum Noise: A Handbook of Markovian and Non-Markovian Quantum Stochastic Methods with Applications to Quantum Optics}},
    year = {2004},
    author = {Gardiner, Crispin and Zoller, Peter},
    publisher = {Springer Science {\&} Business Media},
    isbn = {978-3-540-22301-6}
}

@article{Rossini2023SingleDynamics,
    title = {{Single Qubit Error Mitigation by Simulating Non-Markovian Dynamics}},
    year = {2023},
    author = {Rossini, Mirko and Maile, Dominik and Ankerhold, Joachim and Donvil, Brecht I. C},
    month = {3},
    url = {http://arxiv.org/abs/2303.03268 http://dx.doi.org/10.1103/PhysRevLett.131.110603},
    doi = {10.1103/PhysRevLett.131.110603},
    arxivId = {2303.03268}
}

@article{Kleinekathofer2002StochasticClass,
    title = {{Stochastic unraveling of time-local quantum master equations beyond the Lindblad class}},
    year = {2002},
    journal = {Physical Review E - Statistical Physics, Plasmas, Fluids, and Related Interdisciplinary Topics},
    author = {Kleinekath{\"{o}}fer, Ulrich and Kondov, Ivan and Schreiber, Michael},
    number = {3},
    pages = {1--5},
    volume = {66},
    doi = {10.1103/PhysRevE.66.037701},
    issn = {1063651X},
    pmid = {12366307},
    arxivId = {quant-ph/0208084}
}

@article{Pinol2024TellingAverages,
    title = {{Telling different unravelings apart via nonlinear quantum-trajectory averages}},
    year = {2024},
    journal = {Physical Review Research},
    author = {Pi{\~{n}}ol, Eloy and Mavrogordatos, Th. K. and Keys, Dustin and Veyron, Romain and Sierant, Piotr and Angel Garc{\'{i}}a-March, Miguel and Grandi, Samuele and Mitchell, Morgan W. and Wehr, Jan and Lewenstein, Maciej},
    number = {3},
    month = {9},
    pages = {L032057},
    volume = {6},
    url = {https://link.aps.org/doi/10.1103/PhysRevResearch.6.L032057},
    doi = {10.1103/PhysRevResearch.6.L032057},
    issn = {2643-1564}
}

@article{Gaona-Reyes2025TheoreticalUnravelings,
    title = {{Theoretical limits of protocols for distinguishing different unravelings}},
    year = {2025},
    journal = {Physical Review Research},
    author = {Gaona-Reyes, J. L. and Altamura, D. G. A. and Bassi, A.},
    number = {4},
    month = {12},
    pages = {043295},
    volume = {7},
    url = {https://link.aps.org/doi/10.1103/6qnt-t3wl},
    doi = {10.1103/6qnt-t3wl},
    issn = {2643-1564}
}

@article{Dalton2001TheoryProcesses,
    title = {{Theory of pseudomodes in quantum optical processes}},
    year = {2001},
    journal = {Physical Review A. Atomic, Molecular, and Optical Physics},
    author = {Dalton, B. J. and Barnett, S. M. and Garraway, B. M.},
    number = {5},
    pages = {538131--5381321},
    volume = {64},
    doi = {10.1103/PhysRevA.64.053813},
    issn = {10502947},
    arxivId = {quant-ph/0102142}
}

@article{Smirne2016UltimateEstimation,
    title = {{Ultimate Precision Limits for Noisy Frequency Estimation}},
    year = {2016},
    journal = {Physical Review Letters},
    author = {Smirne, Andrea and Ko{\l}ody{\'{n}}ski, Jan and Huelga, Susana F. and Demkowicz-Dobrza{\'{n}}ski, Rafał},
    number = {12},
    pages = {1--6},
    volume = {116},
    doi = {10.1103/PhysRevLett.116.120801},
    issn = {10797114},
    arxivId = {1511.02708}
}

@article{Donvil2023Unraveling-pairedChannels,
    title = {{Unraveling-paired dynamical maps recover the input of quantum channels}},
    year = {2023},
    journal = {New Journal of Physics},
    author = {Donvil, Brecht and Muratore-Ginanneschi, Paolo},
    number = {5},
    pages = {53031},
    volume = {25},
    publisher = {IOP Publishing},
    url = {https://doi.org/10.1088/1367-2630/acd4dc},
    doi = {10.1088/1367-2630/acd4dc},
    issn = {13672630}
}

@article{Dalibard1992Wave-functionOptics,
    title = {{Wave-function approach to dissipative processes in quantum optics}},
    year = {1992},
    journal = {Physical Review Letters},
    author = {Dalibard, Jean and Castin, Yvan and M{\o}lmer, Klaus},
    number = {5},
    pages = {580--583},
    volume = {68},
    doi = {10.1103/PhysRevLett.68.580},
    issn = {00319007}
}

@misc{github,
    howpublished = {The code used for the simulations is available at \url{https://github.com/federicoSettimo/Review_jumps.git}},
}

@article{Borah2022,
  title = {Measurement-based estimator scheme for continuous quantum error correction},
  author = {Borah, Sangkha and Sarma, Bijita and Kewming, Michael and Quijandr\'{\i}a, Fernando and Milburn, Gerard J. and Twamley, Jason},
  journal = {Phys. Rev. Res.},
  volume = {4},
  issue = {3},
  pages = {033207},
  numpages = {10},
  year = {2022},
  month = {Sep},
  publisher = {American Physical Society},
  doi = {10.1103/PhysRevResearch.4.033207},
  url = {https://link.aps.org/doi/10.1103/PhysRevResearch.4.033207}
}

@misc{Xu2026Review,
   author = {Meng Xu and Vasilii Vadimov and J. T. Stockburger and J. Ankerhold},
   doi = {10.1103/w3nw-hbjc},
   issue = {2002},
   month = {1},
   pages = {1-28},
   title = {Simulating Non-Markovian Dynamics in Open Quantum Systems},
   url = {http://arxiv.org/abs/2601.02160 http://dx.doi.org/10.1103/w3nw-hbjc},
   year = {2026},
    archivePrefix = {arXiv},
    arxivId = {2601.02160},
}

@misc{Settimo-SSE-RO_misc,
   author = {Federico Settimo},
   pages = {1-12},
   title = {A Stochastic Schrödinger Equation for the Generalized Rate Operator Unravelings},
   url = {http://arxiv.org/abs/2507.01107},
   year = {2025}
}

@misc{Donvil2023QuantumMitigation_misc,
   author = {Brecht. I. C Donvil and Rochus Lechler and Joachim Ankerhold and Paolo Muratore-Ginanneschi},
   month = {5},
   title = {Quantum Trajectory Approach to Error Mitigation},
   url = {http://arxiv.org/abs/2305.19874},
   year = {2023}
}

@misc{Mondal2025_misc,
   author = {Sujay Mondal and Siddhartha Dutta and Abhijit Bandyopadhyay},
   month = {10},
   pages = {1-30},
   title = {Tensor-Network-Based Unraveling of Non-Markovian Dynamics in Large Spin Chains via the Influence Martingale Approach},
   url = {http://arxiv.org/abs/2510.11200},
   year = {2025}
}

@misc{Gu2025_misc,
   author = {Xu-Ke Gu and Li-Zhou Tan and Franco Nori and J. Q. You},
   month = {3},
   pages = {1-6},
   title = {Exploring Dynamics of Open Quantum Systems in Naturally Inaccessible Regimes},
   url = {http://arxiv.org/abs/2503.06946},
   year = {2025}
}

\end{document}